\documentclass[longauth,subabstract]{aa}	

\usepackage{graphicx}
\usepackage{array}
\usepackage{lscape}                                
\usepackage{txfonts}
\usepackage{natbib}
\usepackage{longtable}
\usepackage{color}
\usepackage{lscape}

\bibpunct{(}{)}{;}{a}{}{,}

\begin{document}
\title{The VLT-FLAMES Tarantula Survey\thanks{Based on observations at the European Southern 
Observatory Very Large Telescope in program 182.D-0222.}}
\subtitle{XIII: On the nature of O\,Vz stars in 30\,Doradus}
\author{C.~Sab\'in-Sanjuli\'an\inst{1,2}, S.~Sim\'on-D\'iaz\inst{1,2},
  A.~Herrero\inst{1,2}, N.~R.~Walborn\inst{3}, J.~Puls\inst{4},
  J.~Ma\'iz~Apell\'aniz\inst{5}, C.~J.~Evans\inst{6}, 
  I.~Brott\inst{7}, A.~de~Koter\inst{8,9}, M.~Garcia\inst{10},
N.~Markova\inst{11}, F.~Najarro\inst{10}, O.~H.~Ram\'irez-Agudelo\inst{8},
H.~Sana\inst{3}, \\
W.~D.~Taylor\inst{6} and J.~S.~Vink\inst{12}}

\institute{Instituto de Astrof\'isica de Canarias, E-38200 La Laguna,
  Tenerife, Spain  
  \and Departamento de Astrof\'isica, Universidad de La Laguna,
  E-38205 La Laguna, Tenerife, Spain 
  \and Space Telescope Science Institute, 3700 San Martin Drive,
  Baltimore, MD 21218, USA 
  \and Universit\"ats-Sternwarte, Scheinerstrasse 1, 81679 M\"unchen,
  Germany 
  \and Instituto de Astrof\'isica de Andaluc\'ia-CSIC, Glorieta de la
  Astronom\'ia s/n, E-18008 Granada, Spain 
\and UK Astronomy Technology Centre, Royal Observatory Edinburgh,
  Blackford Hill, Edinburgh, EH9 3HJ, UK 
 \and University of Vienna, Department of Astrophysics,
  T\"{u}rkenschanzstr. 17, 1180, Vienna, Austria 
  \and Astronomical Institute Anton Pannekoek, University of
  Amsterdam, Kruislaan 403, 1098 SJ, Amsterdam, The Netherlands 
  \and Instituut voor Sterrenkunde, Universiteit Leuven,
  Celestijnenlaan 200 D, 3001, Leuven, Belgium 
  \and Centro de Astrobiolog\'ia (CSIC-INTA), Ctra. de Torrej\'on a
  Ajalvir km-4, E-28850 Torrej\'on de Ardoz, Madrid, Spain 
    \and Institute of Astronomy with NAO, Bulgarian Academy of Sciences,
  PO Box 136, 4700 Smoljan, Bulgaria 
\and Armagh Observatory, College Hill, Armagh, BT61 9DG, Northern Ireland, UK}%

\offprints{C. Sab\'in-Sanjuli\'an, email: \texttt{cssj@iac.es}}

\date{Submitted/Accepted}

\titlerunning{VFTS: The nature of O\,Vz stars in 30\,Dor}
\authorrunning{Sab\'in-Sanjuli\'an et al.}  

\abstract {O\,Vz stars, a subclass of O-type dwarfs characterized by
  having \ion{He}{ii}\,$\lambda$4686 stronger in absorption than any
  other helium line in their blue-violet spectra, have been suggested
  to be on or near the zero-age-main-sequence (ZAMS). If their youth
  were confirmed, they would be key objects with which to advance our knowledge
  of the physical properties of massive stars in the early stages of
  their lives.}{We test the hypothesis of O\,Vz stars being at a
  different (younger) evolutionary stage than are normal
  O-type dwarfs.}{We have performed the first comprehensive quantitative
  spectroscopic analysis of a statistically-meaningful sample of O\,Vz
  and O\,V\ stars in the same star-forming region, exploiting the large
  number of O\,Vz stars identified by the VLT-FLAMES Tarantula Survey
  in the 30\,Doradus region of the Large Magellanic Cloud (LMC).  We
  obtained the stellar and wind parameters of 38 O\,Vz stars (and a
  control sample of 46 O\,V\ stars) using the {\sc fastwind}\ stellar
  atmosphere code and the IACOB-GBAT, a grid-based tool developed for
  automated quantitative analysis of optical spectra of O stars. In the framework of a differential study, we
  compared the physical and evolutionary properties of both samples,
  locating the stars in the log\,{\em g}\,vs.\,log\,$T_{\rm eff}$,
  log\,{\em Q}\,vs.\,log\,$T_{\rm eff}$, and log\,$L/L_{\odot}$\,vs.\,log\,$T_{\rm eff}$\
  diagrams. We also investigated the predictions of the {\sc fastwind}\ code
  regarding the O\,Vz phenomenon.}  {We find a differential
  distribution of objects in terms of effective temperature, with O\,Vz stars 
  dominant at intermediate values.  The O\,Vz stars in
  30\,Doradus tend to be younger (i.e., closer to the ZAMS) and less
  luminous, and they have weaker winds than the O\,V\ stars, but we also
  find examples with ages of 2-4\,Myr and with luminosities and winds
  that are similar to those of normal O dwarfs.  Moreover, the O\,Vz
  stars do not appear to have higher gravities than the O\,V stars.
  In addition to effective temperature and wind strength, our
  {\sc fastwind}\ predictions indicate how important it is to take other
  stellar parameters (gravity and projected rotational velocity) into
  account for correctly interpreting the O\,Vz phenomenon.}
{In general, the O\,Vz stars appear to be on or very close to the
  ZAMS, but there are some examples where the Vz classification does
  not necessarily imply extreme youth. In particular, the presence of
  O\,Vz stars in our sample at more evolved phases than expected is
  likely a consequence of modest O-star winds owing to the
  low-metallicity environment of the LMC.} 

\maketitle

\section{Introduction}\label{introduction}

One of the open questions in the formation and evolution of massive stars is what is the first stage of their 
lives. A typical high-mass protostar is formed in a molecular cloud, accumulating mass with an accretion 
timescale that could be longer than the contraction timescale \citep[see][]{bernasconi}. This  
would lead to a situation in which a high-mass star begins to burn hydrogen in its core while still 
accreting material from the parental cloud. This zero-age main sequence (ZAMS) star would 
appear embedded and optically obscured in the molecular cloud, even if it is already emitting ultraviolet 
photons and has created a small Str\"omgren sphere around it \citep{churchwell02}. Massive stars may spend 
$\sim$15$\%$ of their lifetime in this embedded stage \citep[][and references therein]{zinnecker07}. 
When the parental cloud is sufficiently dissolved, it is easier to observe the star at optical wavelengths, 
but it is already somewhat evolved. The ZAMS stage of massive stars is therefore very difficult to observe 
at wavelengths where the main diagnostics of stellar parameters are concentrated, i.e., the ultraviolet,
optical, and although to a lesser extent, the near-infrared \citep[see][]{hanson}.

To learn more about the final stage of formation of massive stars, it
is interesting to identify objects that have just passed their
birthline, since these may reveal remnant signatures of the formation
process. Over recent decades, observations of O-type dwarfs in the
Milky Way (MW) and the Magellanic Clouds have opened up optical
studies of O stars apparently on or close to the ZAMS.  \cite{w73}
commented on the presence of O dwarfs in the Tr14 cluster (Carina
Nebula), which displayed higher
\ion{He}{ii}\,$\lambda$4686/\ion{He}{ii}\,$\lambda$4541
absorption-line ratios than in normal dwarfs.  This spectroscopic
feature motivated the definition of a new luminosity subclass, Vz, in
which \ion{He}{ii}\,$\lambda$4686 is deeper than both
\ion{He}{i}\,$\lambda$4471 and \ion{He}{ii}\,$\lambda$4541.

The O\,Vz phenomenon has been suggested to be an `inverse' Of effect.
In Of stars, \ion{He}{ii}\,$\lambda$4686 goes from absorption to
emission when moving from the mildest O((f)) cases to the most extreme
Of examples. This effect is due to the intensity of the stellar wind, which 
increases with luminosity and when the star evolves and departs from
main sequence.  Following this reasoning, the strong
\ion{He}{ii}\,$\lambda$4686 absorption in O\,Vz stars could be
explained by less wind emission filling the line in than in typical
O\,V\ stars. In this scenario, O\,Vz stars are said to be objects
on (or near) the ZAMS, with lower luminosities and weaker winds than
normal O dwarfs \citep[][]{w09}. Therefore, O\,Vz stars would
represent the link between the early phases of star formation (such as
ultra-compact \ion{H}{ii} regions) and the normal, already slightly
evolved, O dwarfs. Moreover, being young and subluminous, they could
also offer a link to the so-called `weak-wind' stars, which are found
to have weaker winds than predicted by the theory of
radiatively-driven winds (\citealt{b03}; \citealt{martins05}). The
reasons for these weak winds are currently under debate but it is
thought that, below a certain luminosity threshold, stars are not able
to initiate their outflows by radiation pressure alone \citep[see][]{lucy,muijres}.

Many stars have been classified as O\,Vz after introduction of the
subclass \citep[viz.][]{morrell91,parker92,w92,w97,parker01}.  
Others have been proposed as belonging to it owing to the presence of 
characteristics that may be linked to the Vz subclass (like the subluminosity or the 
weak winds, see e.g., \citealt{heydari}). We also
refer the reader to the review by \cite{w09}, who presented a
compilation of the 25 O\,Vz stars known at the time\footnote{The
  number of detected/known O\,Vz stars in the MW and the Large
  Magellanic Cloud (LMC) has increased since Walborn's review due to
  large spectroscopic surveys in the MW (GOSSS, \citealt{gosss}; OWN,
  \citealt{own}; IACOB, \citealt{iacob-survey}) and the LMC
  \citep{evans11}.}, located in the Galaxy and Magellanic Clouds.

Despite the potential relevance of O\,Vz stars to the formation and
early evolution of massive stars, only a few have been quantitatively
analyzed to date.  Intringuingly, the limited number of quantitative
results place some on the ZAMS, such as HD 93\,128 \citep{rph04} and
HD 152\,590 \citep{martins05}, while others are seen to depart from the
ZAMS, e.g., HD 42\,088 \citep{martins05} and more recently CPD -58\,2620 and HD 91\,824 (\citealt{nevy13}). 

Three O\,Vz stars in the Small Magellanic Cloud (SMC) were analyzed by
\cite{mokiem06}, who found that two (NGC346-028 and NGC346-051) lay
close to the ZAMS (but, curiously, with enriched helium abundances,
which point to a more evolved stage), while the third (NGC346-031)
appeared older.  These stars have recently been analyzed by \cite{b13}
who, in agreement with \citeauthor{mokiem06}, estimated NGC346-031 to
be older than 2.5\,Myr (but with a lower helium abundance and mass-loss
rate) and placed NGC346-051 close to the ZAMS (younger than 1\,Myr).
However, in contrast to \citeauthor{mokiem06}, they found NGC346-028
to be separated from the ZAMS (at 2-3\,Myr). Interestingly, they
reported an enhanced N/O abundance for NGC346-051 and indicate that,
despite its apparent proximity to the ZAMS, this star could actually
be a more evolved, fast rotator.

A similar puzzling situation was found by \cite{martins12}. Their
analysis of HD 46\,150 and HD 46\,573 found that these two O\,Vz stars
are nitrogen enriched and not particularly closer to the ZAMS than
other O\,V\ stars. However, HD 46\,573 is a known binary
\citep[][]{mason98} and HD 46\,150 is a candidate binary
\citep[][]{mahy09}, which could affect both the spectral
classification and interpretation of the quantitative results.  The
recent case of HD 150\,136 is also interesting as \cite{mahy12} found
that it is part of a triple system, where the primary presents an
incipient P-Cygni profile for \ion{He}{ii}\,$\lambda$4686. After
spectral disentangling, the other two components display O\,Vz
characteristics (although the case of the third component is
uncertain). The position of these two components in the
Hertzsprung-Russell (H--R) diagram is typical of normal O\,V\ stars
(i.e., neither is particularly close to the ZAMS), although their
association with an O3~V primary could be considered an indirect
indication of youth.

These examples serve to illustrate the difficulty of trying to
understand the O\,Vz subclass on a case-by-case basis, not helped by
the paucity of quantitative information available. To clarify their
nature, and their possible links to young (ZAMS) and/or weak-wind
stars, analysis of an extensive and homogeneous dataset is needed.
Such a dataset is now available from the VLT-FLAMES Tarantula Survey
\citep[VFTS,][]{evans11}, an ESO Large Program which has obtained
multi-epoch spectroscopy of over 800 massive stars in the 30\,Doradus
(hereafter 30\,Dor) star-forming region of the LMC.

Spectral classification of the O-type stars observed in the VFTS has
led to the discovery of $\sim$50~O\,Vz stars in 30\,Dor (\citealt{walborn13}). 
In this article we undertake quantitative spectroscopic
analysis of the apparently single O\,Vz stars from the VFTS
\citep[i.e., stars with constant radial velocities from the multi-epoch
spectroscopy, see][]{sana13}, together with a control sample of normal
O\,V\ stars (i.e., without the `z' suffix) from the same survey.  With
these data, we perform a differential study between the O\,Vz and O\,V\ samples 
to test if the O\,Vz stars are (a) younger, (b)
subluminous, (c) of higher gravity, and/or (d) have weaker winds than
normal O\,V\ stars.

In Sect.~\ref{observations} we present the samples of O\,V\ and O\,Vz
stars from the VFTS analyzed in this study.  In Sect.~\ref{method} we
explain the automatic method used to analyze the spectra and note
issues which need to be taken into account; our results are presented
in Sect.~\ref{results}.  Section~\ref{test_fw} presents a theoretical
study of the Vz phenomenon with synthetic {\sc fastwind}\ models, and these
predictions are compared with our results in Sect.~\ref{obs_vs_fw}. We
end with a discussion and our conclusions in Sect.~\ref{conclusion}.

\section{Observations and sample selection}\label{observations}

An introduction to the VFTS, along with an extensive description of
the observing strategy and data reduction was presented by
\cite{evans11}.  In brief, all of the data considered in this study
were obtained using the Medusa mode of the Fibre Large Array
Multi-Element Spectrograph \citep[FLAMES; ][]{pasquini} on the Very
Large Telescope (VLT).  The Medusa fibres couple light from targets
across a 25$'$ field-of-view on the sky into the Giraffe spectrograph,
providing intermediate-resolution spectroscopy of over 130 targets
simultaneously. 

Three of the standard settings of the Giraffe spectrograph were used:
LR02, LR03, and HR15N; the resulting spectral coverage and delivered
resolving power ($R$) of each setting is summarized in
Table~\ref{flames_table}. The same spectra have also been used to
provide detailed spectral classifications (\citealt{walborn13}),
and to investigate the multiplicity \citep{sana13} and rotational
properties \citep{oscar} of the O-type population in 30\,Dor.

\begin{table}[h]
  \caption{Wavelength coverage and resolving power ($R$) of the FLAMES--Giraffe
    settings used in the survey.}
\label{flames_table}
 \centering
\begin{tabular}{ccc}
\hline
\multicolumn{3}{c}{} \\ [-2 ex]

Setting & Range (\AA) & $R$ \\
\hline
\multicolumn{3}{c}{} \\ [-2 ex]
LR02  & 3960-4564 & $\phantom{1}$7\,000\\
LR03  & 4499-5071 & $\phantom{1}$8\,500\\
HR15N & 6442-6817 & 16\,000\\ 
\hline
\end{tabular}
\end{table}

From the $\sim$340 O-type stars in the VFTS, \cite{walborn13}
have identified 48 objects which display \ion{He}{ii}\,$\lambda$4686
absorption that is stronger than any other He line in the blue-violet
region, and hence were classified as O\,Vz.  Ten of these were
identified by \cite{sana13} as showing large-amplitude (i.e., larger
than 20 km\,s$^{-1}$) radial-velocity variations and were discarded from the
sample to avoid possible misinterpretation due to their binary
nature. The final sample of O\,Vz stars considered here therefore
comprises 38 (apparently) single stars. These were complemented with a
control sample of 46 O\,V\ stars from the VFTS, which were selected
following the same criteria, i.e., from the total sample of normal
O\,V\ stars, with those identified as binaries or having uncertain
spectral classifications discarded.

\begin{figure}[t!]
 \centering
  \includegraphics[width=8.5cm]{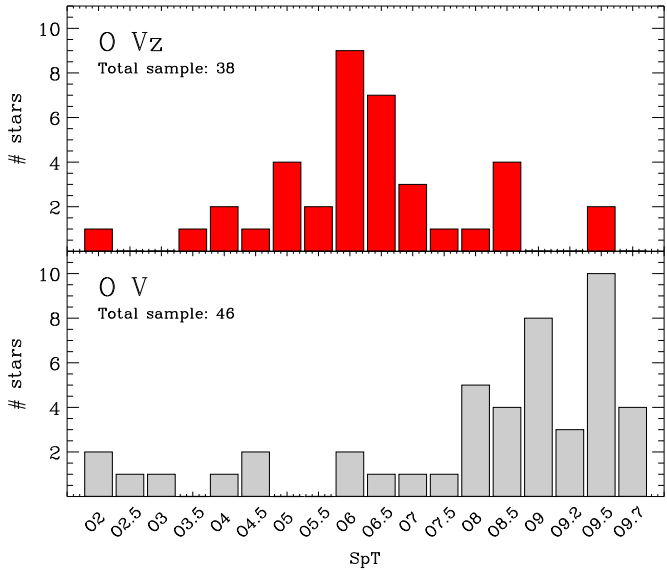}
  \caption{Distribution of spectral types for the O\,Vz (red) and O\,V\
    (gray) stars analyzed.}
 \label{spt_histo}
 \end{figure}

The first three columns of Tables \ref{tab_vz}, \ref{tab_v}, and
\ref{tab_teff} list the VFTS identifications and spectral
classifications (from \citealt{walborn13}) of the stars considered
here.  The spectral types range from O2 to O9.5/O9.7 for both samples
(see Fig.~\ref{spt_histo}). Interestingly, there seems to be a clear
difference in the distribution of spectral types between the samples:
the O\,Vz stars tend to be concentrated at intermediate types
(O5.5\,--\,O7), while the O\,V\ stars mostly have types later than O8.

\begin{figure}[t!]
  \centering
   \includegraphics[width=9 cm]{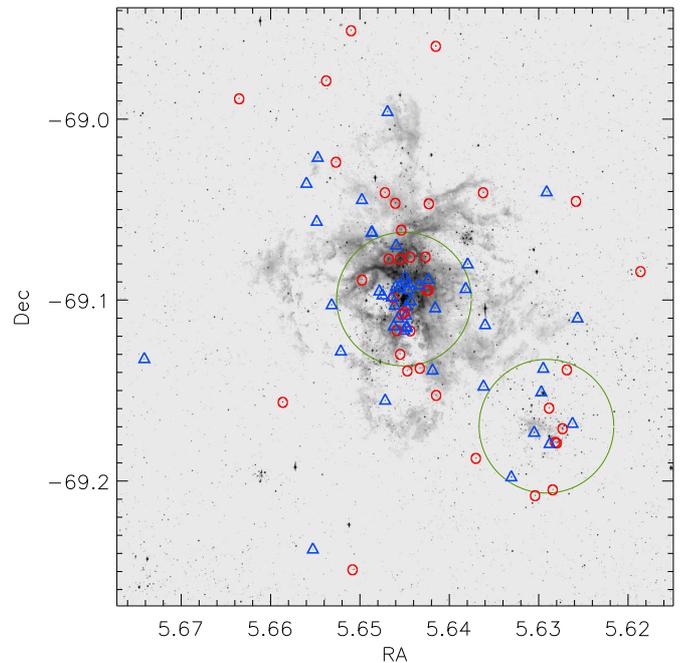}
   \caption{Spatial distribution of the O\,Vz (red circles) and O\,V\
     (blue triangles) stars analyzed in 30 Doradus.  The central and
     south-western green circles indicate the approximate extent of
     the NGC\,2070 and NGC\,2060 clusters, respectively.}
  \label{30dor}
 \end{figure}

Figure~\ref{30dor} shows the spatial distribution of the O\,Vz and O\,V
samples. \cite{walborn13} found that the O\,Vz stars seem to
be concentrated within regions of recent and current star formation
(i.e., the ionizing clusters NGC\,2060 and 2070, as well as an
east-west band at the northern edge of the nebula). \citeauthor{walborn13}
also found that the distribution of O\,Vz stars is quite different
to that of the rapidly-rotating stars. However, when the spatial
distribution of O\,V\ and O\,Vz stars are compared (without any
constraint in the rotational velocities) there does not seem to be a
clear difference between them.
 
 \begin{figure*}[t!]
  \centering
   \includegraphics[scale=0.68,angle=90]{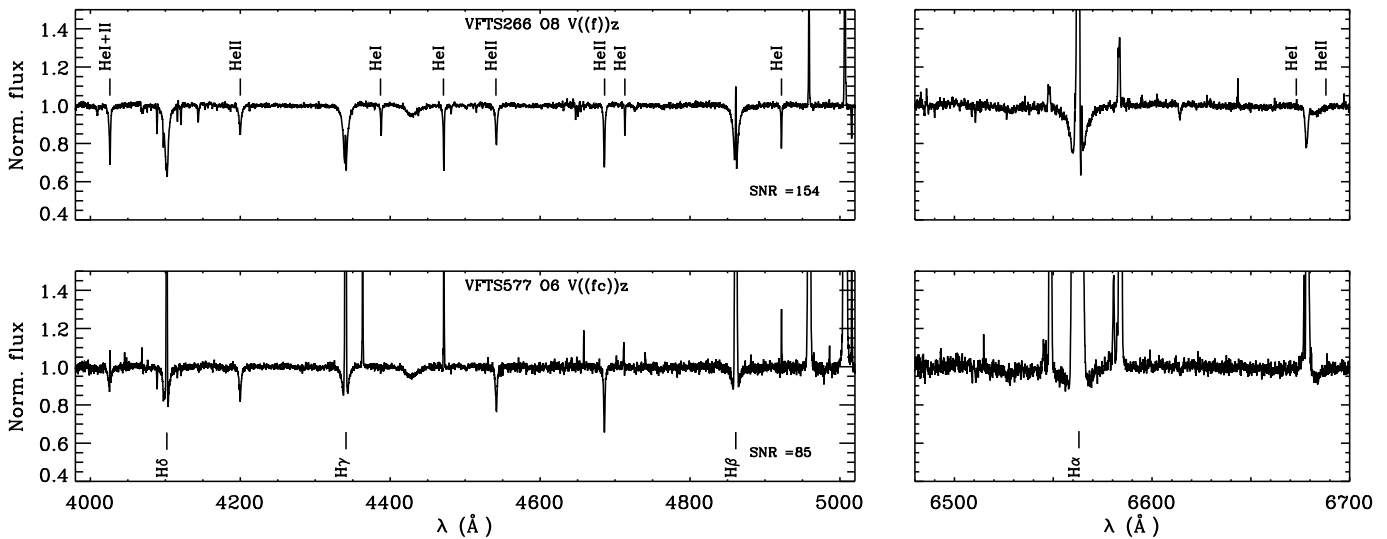}
   \caption{Two examples of the FLAMES--Medusa spectra used in this
     study, in which VFTS\,577 (lower panel) has serious
     contamination by nebular emission.  The diagnostic hydrogen and
     helium lines used in our analysis are indicated (see
     Table~\ref{hhe_lines}).}
  \label{plot_ex}
 \end{figure*}

 Figure~\ref{plot_ex} shows two characteristic Medusa
 spectra\footnote{We refer the reader to \cite{walborn13}
   for further examples of O\,Vz spectra from the VFTS.  Also note
   that the spectra and best-fitting models for the primary diagnostic
   lines used in this study are shown in Figs.~\ref{spec_vz1} to
   \ref{spec_v4}.}, which include suitable diagnostic lines of
 hydrogen and helium for the quantitative analysis presented in this
 paper.  Given that detected binaries were discarded from the samples,
 the multi-epoch spectra for each object were combined (after
 individual normalization) to increase the signal-to-noise (S/N)
 ratio. In addition, the wavelength scales of the combined spectra
 were corrected to the rest frame, using the velocities from
 \cite{sana13}.  The S/N of the combined spectra is typically
 $\sim$150 (with 27 cases with S/N\,$>$\,200, and 14 with
 S/N\,$<$\,100).  As illustrated in Fig.~\ref{plot_ex}, nebular lines
 emitted by the ionized gas are present in all the spectra,
 contaminating the stellar absorption lines (mainly the hydrogen
 Balmer and \ion{He}{i} lines) to differing degrees. This has
 important consequences for the accurate determination of the stellar
 and wind parameters of some of the stars in our sample (see
 Sect.~\ref{analisis}), especially in cases where the nebular emission
 lines are strong and broad.

\section{Quantitative spectroscopic analysis}\label{method}
\label{analisis}

The stellar parameters of our sample were determined using standard
techniques \citep[see e.g.,][]{h92,h02,rph04}, employing the
{\sc fastwind}\, stellar atmosphere code \citep{santolaya,puls05} and the
IACOB-Grid Based Automatic Tool (IACOB-GBAT), a tool developed for
automated spectroscopic analysis of O stars.

The tool uses a large grid of {\sc fastwind}\, models which span a wide
range of stellar and wind parameters (which are optimized for the
analysis of O-type stars), and uses a $\chi^{2}$ algorithm to perform
comparisons between the observed and synthetic \ion{H}{i},
\ion{He}{i}, and \ion{He}{ii} line profiles; a detailed description of
the IACOB-GBAT was given by \cite{iacob11}. In this section we comment
on some of the details concerning application of the IACOB-GBAT to
analysis of the VFTS data.

\subsection{Grid of {\sc fastwind} models}\label{grid}

\begin{table}[t!]
  \caption{Parameter ranges considered in the {\sc fastwind}\ grid incorporated in the 
    IACOB-GBAT used in this study.}
\label{grid_param}
 \centering
\begin{tabular}{ll}
\hline
\multicolumn{2}{c}{} \\   [-2 ex]
Parameter & Ranges and/or specific values \\
\hline
\multicolumn{2}{c}{} \\   [-2 ex]
$Z$ & 0.5 $Z_{\odot}$ \\
$T_{\rm eff}$ & $\ge$\,25000 K \, \, [Step: 1000 K] \\
log\,{\em g}& [2.6 - 4.3]\,dex \, \, [Step: 0.1\,dex]\\
Y(He)\,$ ^{(1)}$ & 0.06, 0.09, 0.10, 0.12, 0.15, 0.20, 0.25, 0.30\\
$\zeta_{\rm t}\,^{(2)}$ & 5, 10, 15, 20 km\,s$^{-1}$\ \\
log\,Q & --15.0, --14.0, --13.5, --13.0, --12.7, --12.5, --12.3, --12.1\\
       & --11.9, --11.7\\
$\beta\,^{(2)}$ & 0.8, 1.0, 1.2, 1.5, 1.8 \\
\hline
\multicolumn{2}{c}{} \\   [-2 ex]
\multicolumn{2}{l}{$^{(1)}$Defined as Y(He)\,=\,$\rm{N(He})/N(\rm{H})$.}\\
\multicolumn{2}{l}{$^{(2)}$See, however, notes in Sections \ref{beta} and \ref{micro}. } \\
\end{tabular}
\end{table}

From the various {\sc fastwind}\ grids which are presently incorporated in
the IACOB-GBAT, we have considered the models computed for
Z\,=\,0.5\,Z$_{\odot}$, corresponding to the approximate metallicity
of the LMC \citep[e.g.,][]{mokiem07}.  The 0.5\,Z$_{\odot}$ grid was
constructed assuming six free parameters: effective temperature 
$T_{\rm eff}$, logarithmic surface gravity log\,{\em g}, helium abundance by number
relative to hydrogen Y(He), microturbulence $\xi_{\rm t}$, the
exponent of the wind velocity law $\beta$, and the wind-strength
parameter $Q$. This parameter was defined by \cite{p96}, and it groups the mass-loss rate $\dot{M}$,
the terminal velocity $v_{\infty}$, and the stellar radius $R$ together
under the optical-depth invariant
$Q$\,=\,$\dot{M}$\,$(R\,v_{\infty})^{-3/2}$.  
Depending on $\rho^2$ processes
in O and B stars profiles of H$\alpha$ and other lines are nearly identical for the same Q. 
Therefore, this is the parameter actually covered by the analysis. Mass-loss rates can only be determined 
if $v_{\infty}$ is known (and radius, but this is obtained in our analysis). 
Unfortunately, $v_{\infty}$ is not known for our stars. Although we could adopt a calibration \citep[see e.g.,][]{vink2001} 
we refrain from it for two reasons: a) it would introduce an unknown error in the mass-loss rate and b) we do not know a priori 
whether there is a difference in the $v_{\infty}$ of O\,V and O\,Vz stars. Therefore, we stick to Q for the analysis, as this is 
the parameter derived directly from the spectroscopic analysis.
$v_{\infty}$ would require UV observations
of the targets, which is much more expensive in observing time given the limited multiplexing capabilities of the HST.

Our grid of models were computed under the assumption of no clumping. 
As shown by \cite{najarro11}, model fits for stars with thin
winds (like O\,V stars in the LMC) require no clumping and therefore, to first order, are probably unclumped. 
Moreover, as this is a differential analysis, only the
difference in behavior of this already weak clumping between O\,V and O\,Vz would 
be relevant. Hence, we consider the use of unclumped models in the study presented here fully justified.
A summary of the parameter
ranges covered by the grid is given in Table~\ref{grid_param}.

\subsection{Line broadening parameters}

As part of the VFTS series of papers, \cite{oscar} have provided
estimates for the projected rotational velocities ($v$\,sin\,$i$) for the
single O-type stars from the survey. Given the objectives and scale
of their study, they did not investigate the effects of
macroturbulent broadening in their sample, which may be important
in the context of the analysis presented here.

We therefore proceeded as follows: we initially used the $v$\,sin\,$i$\
values from \citeauthor{oscar} and checked if the global broadening of
the \ion{He}{i-ii} lines was properly reproduced. In cases with a
clear disagreement between the global broadening of the observed and
synthetic lines from the best-fitting model, we iterated on the
$v$\,sin\,$i$\ value to improve the final fit (in this way, we mimic the effects of macroturbulence).

In addition, we needed to modify the $v$\,sin\,$i$\ values for 17 of the 84 stars analyzed
here.  Six of these were stars with low $v$\,sin\,$i$\ values for which, as
indicated by \citeauthor{oscar}, the estimates were less certain given
the adopted methods. A further nine cases correspond to stars with
$v$\,sin\,$i$\,$\ge$\,150\,km\,s$^{-1}$\ with weak or heavily contaminated
\ion{He}{i} lines. Finally, we detected two stars with much broader
\ion{He}{i} lines than \ion{He}{ii} lines; both stars have close
companions in images from the {\em Hubble Space Telescope (HST)},
suggesting that the spectra are composite (see Sect.~\ref{results}).

\subsection{Diagnostic lines}\label{diagnostic}

\begin{table}[tt!]
  \caption{Hydrogen and helium lines used in the spectroscopic analysis (from the three FLAMES--Medusa settings).}
\label{hhe_lines}
 \centering
\begin{tabular}{lll}
\hline
\multicolumn{3}{c}{}  \\ [-2 ex]
LR02 & LR03 & HR15N \\
\hline
\multicolumn{3}{c}{}  \\ [-2 ex]

H$_{\delta}$\,(4102) & H$_{\beta}$\,(4861) & H$_{\alpha}$\,(6563) \\
H$_{\gamma}$\,(4341) & \ion{He}{i}\,4713   & \ion{He}{i}\,6678 \\
\ion{He}{i+ii}\,4026 & \ion{He}{i}\,4922   & \ion{He}{ii}\,6683 \\
\ion{He}{i}\,4387    & \ion{He}{ii}\,$\lambda$4541  & \\
\ion{He}{i}\,$\lambda$4471    & \ion{He}{ii}\,$\lambda$4686  & \\
\ion{He}{ii}\,4200   & \\
\hline
\end{tabular}
\end{table}

The diagnostic lines used for the stellar parameter determination of
our samples of O\,Vz and O\,V stars are summarized in
Table~\ref{hhe_lines}. The same weight was initially given to each
line, but the full \ion{H}{i} and \ion{He}{i} profiles could not be
used in many of the stars due to nebular contamination. Note that the
fitting strategy followed by the IACOB-GBAT requires all these
contaminated regions within the line profiles to be clipped before the
analysis.

Nebular hydrogen Balmer lines are present in all the spectra, with
H$_{\alpha}$ the most affected line (see
Figs.~\ref{spec_vz1}-\ref{spec_v3}, available online). In all cases,
while the nebular emission contaminates the cores of H$_{\beta}$,
H$_{\gamma}$, and H$_{\delta}$ (to differing extents), the wings of
these lines can still provide reliable diagnostics of the stellar
gravity.  The situation for the H$_{\alpha}$ line is worse; there is
an important fraction of stars for which the whole profile is heavily
contaminated. This could impose a challenge in the determination of
the wind-strength Q-parameter but, in the case of O-dwarfs (where
changes in the wind properties only affect the core of H$_{\alpha}$),
this issue is mitigated by the inclusion of
\ion{He}{ii}\,$\lambda$4686 in the analysis. As with H$_{\alpha}$,
\ion{He}{ii}\,$\lambda$4686 has a strong dependence on the
Q-parameter, but is not affected by nebular contamination in our data.

The \ion{He}{i} lines are contaminated by nebular emission (to
different degrees) in $\sim$70\% of the analyzed sample.  This is not
a significant problem for mid/late O-type stars with $v$\,sin\,$i$\ above a
certain value (where the \ion{He}{i} lines are strong and broad), but
the accuracy (and even reliability) of the analyzes can be affected
for narrow-lined and/or early-type stars.  Thus, results from the
IACOB-GBAT analysis of the stars in which important parts of the
\ion{He}{i} line profiles were clipped must be handled with care and
were carefully checked.  As described in the next section, the stellar
parameter which can be most affected by nebular contamination of the
\ion{He}{i} lines is $T_{\rm eff}$, which also produces a secondary effect on
log\,{\em g}\ and Y(He).

\subsection{Stellar parameter determination}\label{parameter-method}

The strategy followed by the IACOB-GBAT for the determination of the
stellar parameters is based on the minimization of the quantity
$\chi^2_T$ \citep[as described by][]{iacob11}, which is a measurement of
the global goodness-of-fit of a given synthetic spectrum to the
observed data. The investigation of how this quantity varies with the
different stellar and wind parameters of interest (i.e.,  $\chi^2_T$
distributions) provides an objective and homogeneous determination of
the best values and associated uncertainties of the parameters under
study, and also identifies cases where a given parameter cannot be
properly constrained (or if only upper/lower limits can be
determined).

Before discussing the general strategy followed for the cases of $T_{\rm eff}$, log\,{\em g}, Y(He), and
log\,{\em Q}, we present below some notes on the results for the $\beta$\,--\,parameter and the
microturbulence.

\subsubsection{$\beta$\,--\,parameter}\label{beta}

The {\sc fastwind}\ grid was computed for six values of the
$\beta$\,--\,parameter, but initial tests with $\beta$ as a free
parameter indicated that the $X^2_{\rm T}$ distributions were
degenerate, i.e., $\beta$ cannot be constrained for the stars analyzed
in this paper. Therefore, we adopted $\beta$\,$=$\,0.8, a typical
value for O-type dwarfs \citep[see, e.g.,][]{rph04}, and predicted by
theory including a finite disk and multiple scattering
\citep[][]{mueller08}. This approach reduces the computational time
required for the analysis of each star down to about 5\,--\,20\,min,
depending on the specific case.

\subsubsection{Microturbulence}\label{micro}

We also found that the $\chi^2_{\rm T}$ distribution for microturbulence
is degenerate in many cases, with others indicating a somewhat smaller
$\chi^2_{\rm T}$ (i.e., better fit) for a microturbulence of 5\,km\,s$^{-1}$. We
therefore concluded that microturbulence cannot be adequately constrained
for the stars in this study and adopted $\xi_{\rm t}$\,=\,5\,km\,s$^{-1}$\ in our
analysis. While the value may be higher in the case of mid and early O
dwarfs, consideration of other values would only vary the final
parameters within the standard error box \citep[see
also][]{villamariz}.
 
\subsubsection{$T_{\rm eff}$, log\,{\em g}, Y(He), and log\,{\em Q}}

  \begin{figure}[t!]
  \centering

   \includegraphics[scale=0.45]{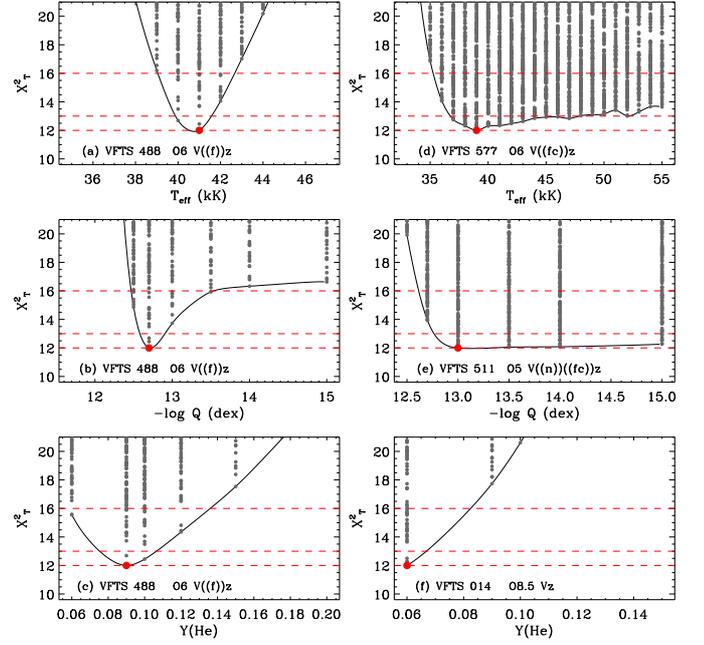}  
   \caption{\label{degen} Examples of $\chi_T^2$ distributions from the
     IACOB-GBAT analysis. From top to bottom, $T_{\rm eff}$, log\,{\em Q}\ and Y(He).
     Horizontal red dashed lines correspond to the minimum $\chi^2_{\rm
       T}$ value and the associated $1\sigma$ and $2\sigma$ deviating
     limits from the best-fitting model (computed as $\chi^2_{\rm T, min}$+1 and $\chi^2_{\rm T,min}$+4, respectively). Black 
curves result from a smooth fit to the lower envelope of each distribution.}
 \end{figure}

In general, we were able to provide estimates (plus associated formal errors) for these four parameters for
most of the analyzed stars. However, we also found some cases in which we could only provide upper/lower 
limits, and detected a few stars for which a quantitative spectroscopic analysis based exclusively on the H 
and He lines could not provide reliable estimates of the effective temperature. Some notes on how
we dealt with these situations are described below.

Figure~\ref{degen} shows some illustrative examples of the $\chi^2_{\rm T}$ distributions resulting from the 
IACOB-GBAT analysis of stars in our sample. Panels (a) to (c) can be considered as typical 
examples of distributions from which the associated 
parameter could be properly determined. The distributions are symmetric and clearly peaked around a 
central value. In a situation like this, we use the fitting curve of the lower envelope of 
the distribution to provide the central value and associated 1$\sigma$ uncertainties.

For almost half of the stars in the sample we obtained a $\chi^2_{\rm T}$
distribution in log\,{\em Q}\ similar to the one presented in Fig.~\ref{degen}e.
This is mainly a result of the fact that below a certain value of the
mass-loss rate, both H$_{\alpha}$ and \ion{He}{ii}\,$\lambda$4686
become insensitive to changes in the Q-parameter.  This effect is
aggravated by the presence of nebular lines contaminating the core of
the H$_{\alpha}$ line. In such cases, we could provide upper limits
for log\,{\em Q}, given by the intersection of the fitting curve of the lower
envelope of the distribution with the 1$\sigma$ (horizontal) limits. 

We also proceeded in a similar way in situations such as the example
presented in Fig.~\ref{degen}f, where the minimum of the distribution
is found at the grid boundary. This only occurred in a few
cases, for which we could only provide upper limits for the helium
abundance. We suggest that the low abundances found in these cases (4 O\,Vz and 9 O\,V) 
could be a consequence of either the presence of undetected companions that 
would dilute the He lines in the global spectrum (resulting in lower Y(He) abundances than expected), or 
the limitations in the analysis because of low S/N, strong nebular contamination or even 
weak \ion{He}{i} lines.

Finally, Fig.~\ref{degen}d shows a distribution that questions the
possibility to provide a reliable estimation of $T_{\rm eff}$\ for this
specific case. While there is a clear minimum in the distribution, the
degeneracy is quite remarkable below the 2$\sigma$ level. Even if only
those models with value of $\chi^2_{\rm T}$ below the 1$\sigma$ limit are
considered, an uncertainty in $T_{\rm eff}$\ of $\sim$\,5000 K is obtained
following the strategy described above.  This situation mainly occurs
for stars in which an important fraction of the \ion{He}{i} line
profiles are contaminated by nebular emission and/or the \ion{He}{i}
lines are weak/absent. In such cases, the \ion{He}{i}/\ion{He}{ii}
ionization balance (the main diagnostic of $T_{\rm eff}$) is difficult to
determine, and only very rough temperature estimates could be
obtained.  As a consequence, results for the other three parameters
(especially log\,{\em g}\ and Y(He)) must be handled with care.

As mentioned in Sect.~\ref{diagnostic}, roughly 70\% of the stars in
the analyzed sample present some degree of nebular contamination of
the \ion{He}{i} lines (see Figs.~\ref{spec_vz1}\,--\,\ref{spec_v3},
available online).  After checking each case, we found that the
situation is not critical for most objects, and the only consequence
is larger uncertainties. However, for about a dozen stars, mainly
concentrated at the earliest spectral types, it was not possible to
provide robust stellar parameters based exclusively on a HHe analysis
(these are discussed further below).

\section{Results}\label{results}

Results from the quantitative analysis of our sample, based on their H
and He profiles, are presented in Tables \ref{tab_vz} (O\,Vz) and
\ref{tab_v} (O\,V). As indicated above, there are a few objects where
the reliability of the $T_{\rm eff}$\ estimates was doubtful as sufficient
\ion{He}{i} lines were not available for analysis. These stars are
excluded from Tables \ref{tab_vz} and \ref{tab_v}, and will be
analyzed separately using an approach similar to that described by
\cite{rivero12}, which employs \ion{N}{iii-v} lines as the primary
$T_{\rm eff}$\ diagnostic. Complete results of these HHeN analyzes, including
estimates of nitrogen abundances, will be presented by Sim\'on-D\'iaz
et al. (in prep.). For completeness, for these 13 stars, we provide
first estimates of the stellar and wind parameters of interest for
this study in Table~\ref{tab_teff}.

The column entries in Tables~\ref{tab_vz} and \ref{tab_v} are as
follows: (1) VFTS identifier; (2,\,3) Spectral classification from
\cite{walborn13}; (4) absolute visual magnitude $M_V$,
computed from the extinction-corrected, apparent $V$ magnitudes from
Ma\'iz Apell\'aniz et al. (in prep.) and for a distance modulus of
18.5 \citep{gibson}; (5) $v$\,sin\,$i$\ considered in the analysis; (6-13)
derived effective temperature, gravity, helium abundance and
wind-strength Q-parameter, and their estimated uncertainties (more
specifically, the formal errors);
(14) stellar luminosity, obtained from $M_V$ and following the
procedure outlined by \cite{kp2000} and \cite{h92}; (15) comments from
\citeauthor{walborn13} regarding possible binarity/multiplicity.  The entries
in Table~\ref{tab_teff} are similar, but without the formal
uncertainties given the preliminary nature of the results.

As indicated in Table~\ref{tab_vz}, we have adopted a minimum value of
0.1\,dex for the log\,{\em g}\ formal errors, since we consider that
uncertainties below this value are not realistic. There are several
error sources to take into account besides those coming from the
systematics of the method, such as the continuum renormalization
\cite[see, e.g.,][]{berto}.

These tables are complemented with a series of figures
(Figs.~\ref{spec_vz1}\,-\,\ref{spec_v4}, available online) in which
the best-fitting synthetic spectra are overplotted on the observed
data for the most important diagnostic lines for this study, i.e., 
H$_{\gamma}$, \ion{He}{i}\,$\lambda$4471, \ion{He}{ii}\,$\lambda$4541,
\ion{He}{ii}\,$\lambda$4686, and H$_{\alpha}$.

These results have been used to investigate the hypotheses introduced
in Sect.~\ref{introduction} regarding the nature of the O\,Vz stars. We especially emphasize 
that this investigation is based on a \textit{differential analysis comparing results from a homogeneous analysis of 
the O\,Vz and O\,V\ objects within our sample}.
To inform our discussion, we have plotted our data in three different
diagrams: log\,{\em g}\,vs.\,$T_{\rm eff}$, log\,{\em Q}\,vs.\,$T_{\rm eff}$\ and the H--R diagram
(i.e.,  $\log{L/L_{\odot}}$\,vs.\,$T_{\rm eff}$), as shown in Figs.~\ref{gtd},
\ref{diagrama_q} and \ref{hrd}, respectively.  In these figures, the
red and black symbols represent the O\,Vz and O\,V\ stars,
respectively.  Stars with results from the HHeN analysis (i.e.,  from
Table~\ref{tab_teff}) are plotted with diamonds, and the open squares
in Figs.~\ref{gtd} and \ref{hrd} correspond to the stars with close
visual companions.  Finally, the stars for which only an upper limit
in log\,{\em Q}\ could be established (see notes in
Sect.~\ref{parameter-method}) are plotted in Fig.~\ref{diagrama_q} as
open circles, with associated downward arrows.

To compare with the predictions of the evolutionary models, in
Figs.~\ref{gtd} and \ref{hrd} we also plot evolutionary tracks (green
solid lines), the ZAMS (black solid line), and isochrones (gray dotted
lines, for ages up to 6\,Myr) from \cite{brott}, calculated for the
metallicity of the LMC
and an initial rotational velocity\footnote{Selected from the
  model grid of initial rotational velocities as the closest to the
  average $v$\,sin\,$i$\ in our sample.} of 171\,km\,s$^{-1}$.

At this point, we note that while performing the HHeN analysis of the
three O2 stars in our sample (VFTS\,468, 506, and 621), we found that
their Medusa spectra display weak (but clearly detectable) \ion{He}{i}
absorption. The {\sc fastwind}\ models do not predict the detection of
these lines for the effective temperatures derived from the
\ion{N}{iv}/\ion{N}{v} ionization balance. This result, together with
the clear detection of multiple visual components within the 1.2\arcsec\
Medusa fibres for two of them, warns us about the interpretation of
results for these stars as single objects.  These three stars were
therefore omitted from the figures and are not discussed further.
Also omitted are five stars for which photometric data were not
available (indicated in Tables \ref{tab_vz}, \ref{tab_v} and
\ref{tab_teff} by a dash in the $M_V$ and $\log{L/L_{\odot}}$
columns).

\subsection{Effective temperatures and gravities}\label{tyg}

 \begin{figure}[t!]
 \centering
  \includegraphics[width=9.0 cm]{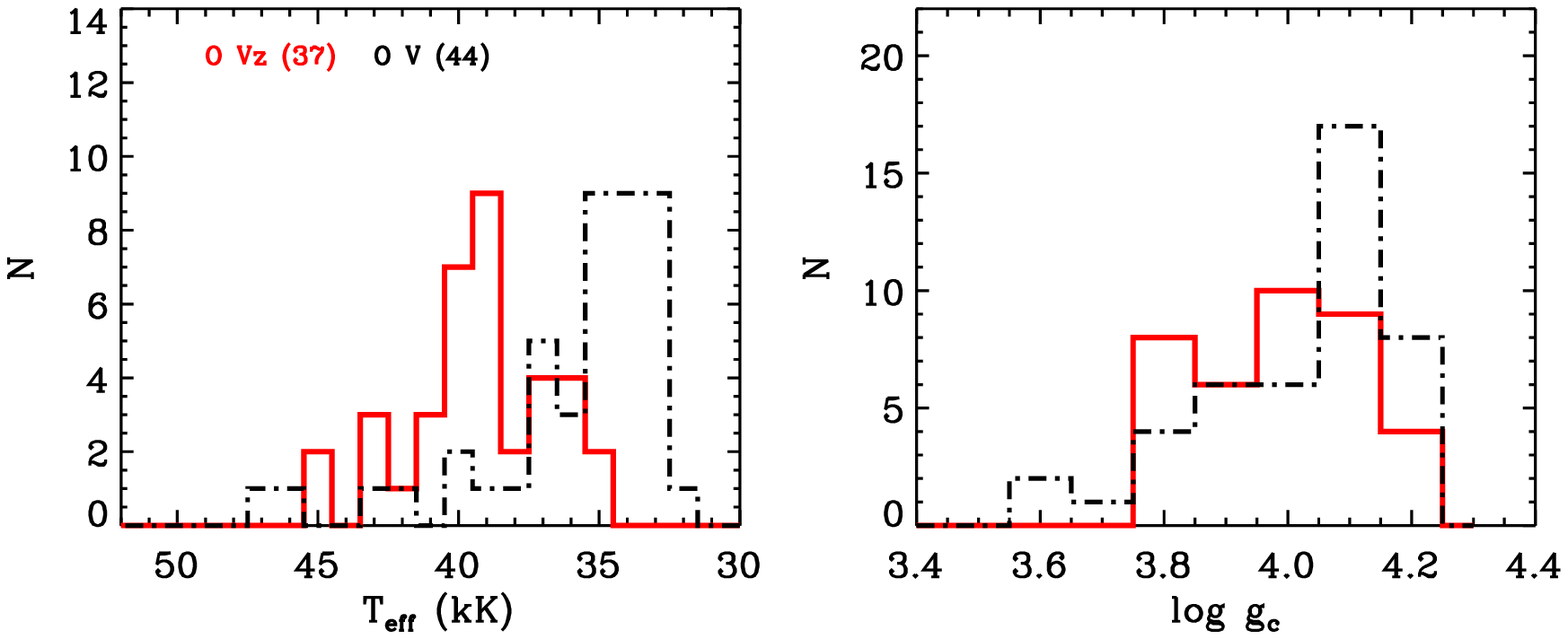}
  \caption{\label{histo_param} Distribution of effective temperatures
    and gravities (corrected for rotation) for the O\,Vz and O\,V\
    stars (red and black dot-dashed lines, respectively).  The three
    O2-type stars discussed in Sect.~\ref{results} are not included
    due to their likely composite nature.}
 \end{figure}

  \begin{figure*}[t!]
 \begin{center}
  \includegraphics[angle=90,width=15cm,trim=0mm 0mm 0mm 0mm,clip]{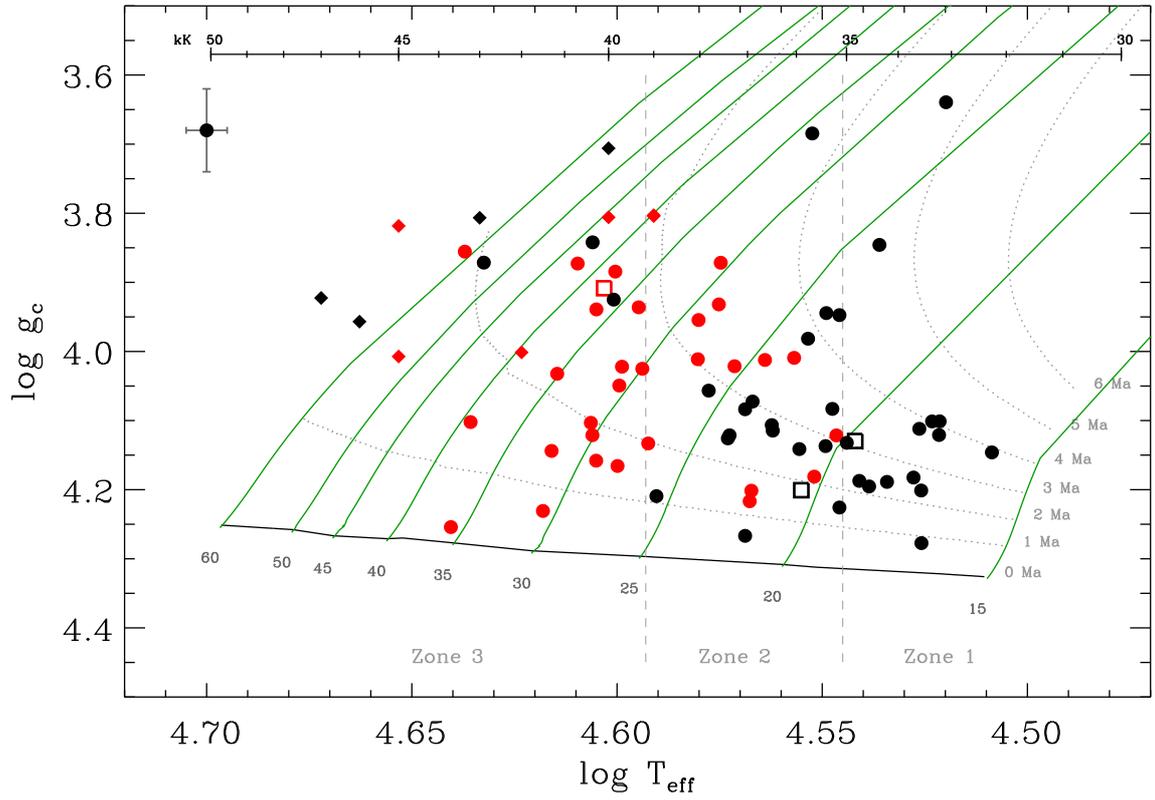}
  \caption{ \label{gtd} log\,{\em g}\,vs.\,log\,$T_{\rm eff}$\ results for the O\,Vz
    and O\,V\ samples (red and black symbols, respectively). Objects with
    confirmed close companions in the Medusa fibres (see footnote to
    Table~\ref{tab_vz}) are plotted as open squares. Stars analyzed
    using nitrogen lines are indicated with diamonds.  To aid the
    discussion, the diagram has been divided into three temperature
    zones (see text for details). }
 \end{center}
 \end{figure*}

  \begin{figure*}[t!!!!]
 \begin{center}
  \includegraphics[angle=90,width=15cm]{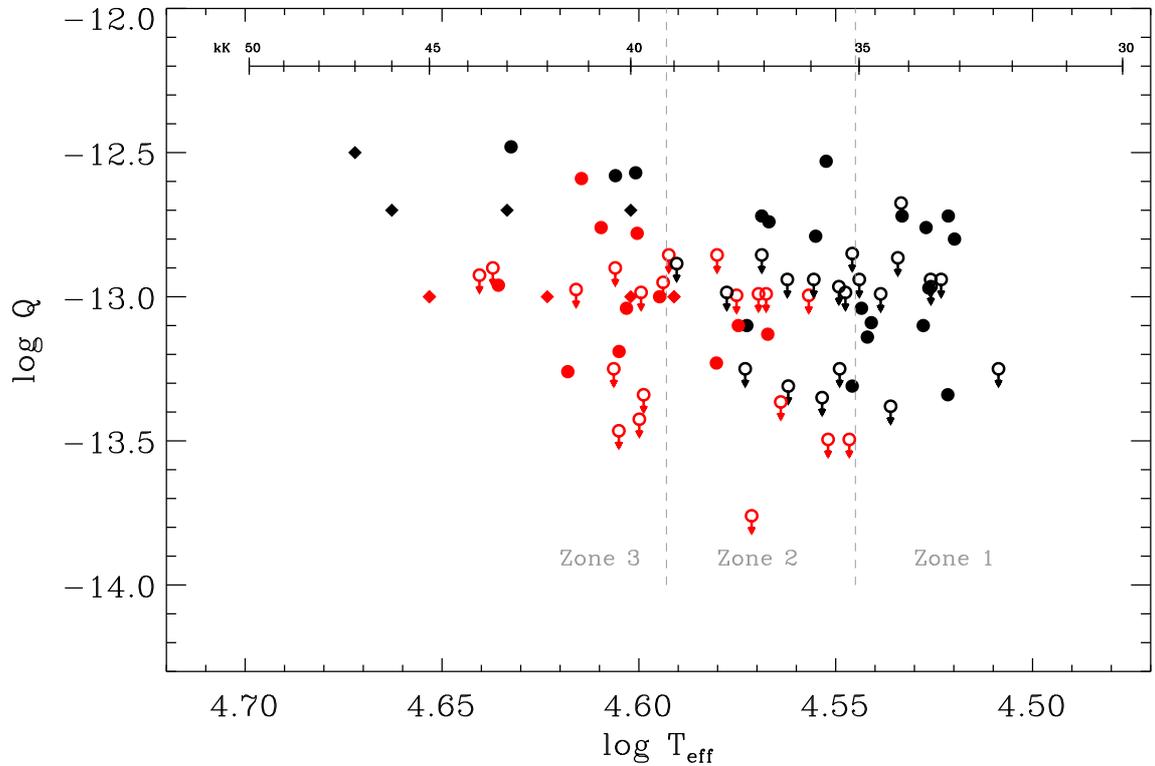}
  \caption{\label{diagrama_q} log\,{\em Q}\,vs.\,log\,$T_{\rm eff}$\ results for the O\,V
    z and O\,V\ samples (red and black symbols, respectively). Cases
    where there was degeneracy in the $\chi_T^2$ distributions used to
    determine log\,{\em Q}\ are indicated as upper limits (with downward
    arrows).  Stars analyzed using nitrogen lines are indicated with
    diamonds. As in Fig.~\ref{gtd}, the diagram has been divided into
    three temperature zones.}
 \end{center}
 \end{figure*}
 
We start by considering three of the stellar parameters which are
directly obtained from the spectroscopic analysis, namely $T_{\rm eff}$,
log\,{\em g}, and log\,{\em Q}. Here we concentrate on the first two, while the
wind properties of the analyzed stars are discussed in next section.
Figure~\ref{histo_param} shows the general distribution of O\,V and O\,Vz 
stars in these two
parameters. As expected from their distribution in spectral types
(see Fig.~\ref{spt_histo}), \textit{the two samples present a
  distinct distribution in temperature, with O\,Vz and O\,V stars
  dominating at intermediate and low temperatures, respectively.} The
two distributions overlap in a transition region spanning
35000\,$\lesssim$\,$T_{\rm eff}$\,$\lesssim$\,40000\,K.  A small but similar
number of objects of both types are found with
$T_{\rm eff}$\,$\ge$\,45000\,K.  The peculiar relative $T_{\rm eff}$\ distributions
of the Vz and V stars suggests that effective temperature plays an
important role in the Vz phenomenon.

Though intriguing, this is not necessarily incompatible with the
postulated nature of the O\,Vz stars. On the contrary, the comparison
of the distributions in gravity challenges the statement that these
objects should have higher gravities than normal O dwarfs due to their
hypothesized proximity to the ZAMS. The range of gravities covered by
both types of objects is similar.  Moreover, contrary to expectations,
the distribution is almost flat for the O\,Vz stars, but peaked
towards higher gravities for the O\,V stars.  Thus, \textit{O\,Vz
  stars do not especially have higher gravities when compared to
  normal O dwarfs.}

Figure~\ref{gtd} compares the distribution of our results in the
log\,{\em g}\,vs.\,log\,$T_{\rm eff}$\ plane. At first inspection, the figure
confirms the two results presented above, but also shows that the O\,V
z stars do not appear particularly closer to the ZAMS (in terms of
gravity) than the normal O dwarfs. This distribution cannot be
reconciled with the expectations, even taking into account the
uncertainties in the derived gravities (up to 0.2\,dex in
a few cases, but with a median value of $\sim$\,0.13\,dex).

\begin{figure*}[t!!!!]
 \begin{center}
  \includegraphics[angle=90,width=15cm,trim=0mm 0mm 0mm 3mm,clip]{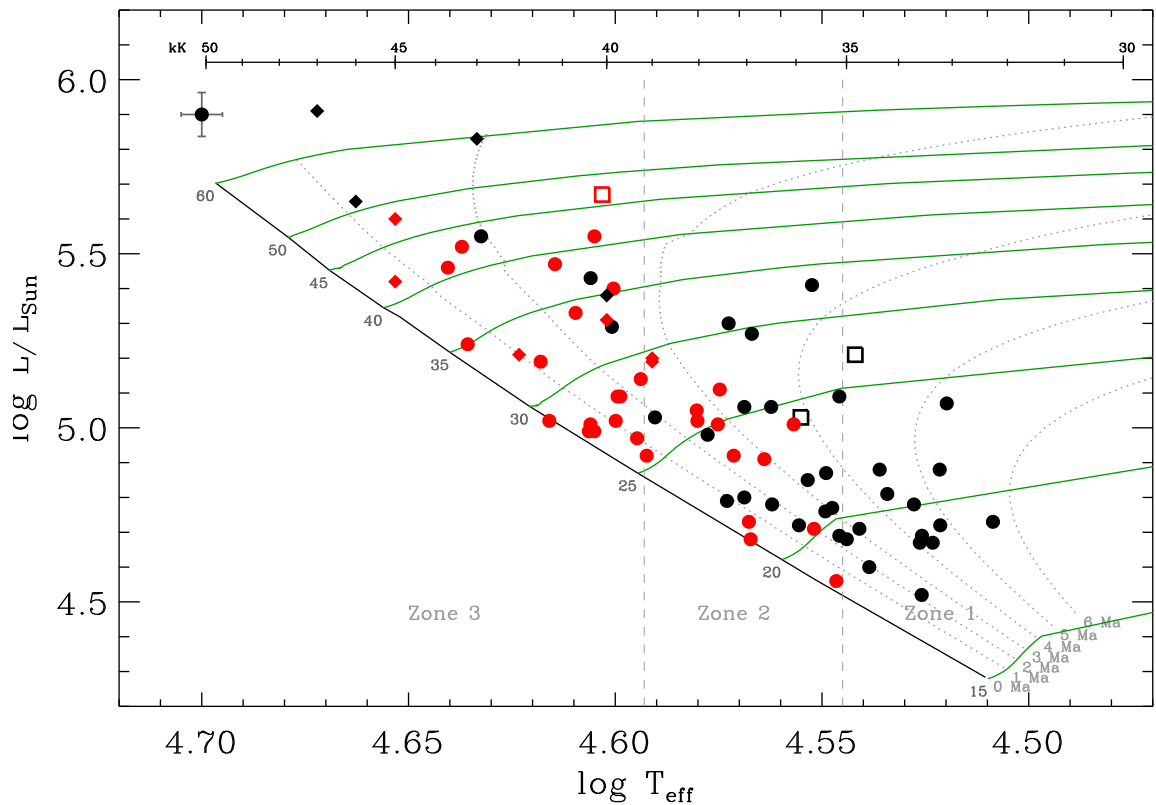}
  \caption{ \label{hrd} Hertzsprung-Russell diagram for O\,Vz and O\,V\
    samples (red and black symbols, respectively). The other symbols
    are the same as those used in Fig.~\ref{gtd}, together with the
    same three temperature zones.} \end{center}
 \end{figure*}

Interestingly, three temperature zones can be distinguished in our
results (illustrative limits between these are indicated by the
vertical dashed lines in Figs.~\ref{gtd} to \ref{hrd}):
\begin{itemize}
 \item Zone 1 ($T_{\rm eff}$\,$\le$\,35000 K), where only O\,V\ stars are found. 
 \item Zone 2 (40000\,$\le$\,$T_{\rm eff}$\,$\le$\,35000 K), where there are similar numbers of O\,Vz and 
 O\,V\ stars. Although both types of objects cover a similar range in gravity, O\,Vz stars tend
 to concentrate at lower gravities and higher effective temperatures than normal O dwarfs.
 \item Zone 3 ($T_{\rm eff}$\,$\ge$\,40000 K), where the number of O\,Vz stars dominates. Only a few O\,V\
  stars are found at relatively low gravities (3.9\,--\,3.7\,dex), while the range of gravities covered by 
 O\,Vz stars ranges from 4.2 to 3.8\,dex.
\end{itemize}

The unexpected distribution of stars in Fig.~\ref{gtd} will be
investigated further in Sect.~\ref{test_fw}, taking into account
{\sc fastwind}\ predictions concerning the relative behavior of the three
main diagnostic lines involved the Vz classification (i.e.,
\ion{He}{i}\,$\lambda$4471, \ion{He}{ii}\,$\lambda$4541, and
\ion{He}{ii}\,$\lambda$4686).

As an aside, we note the recent warning from \cite{massey13} about gravities derived 
using {\sc fastwind}\ models (v10.1). \citeauthor{massey13} presented a comparison of \textsc{cmfgen} \citep[][]{hm98} and 
{\sc fastwind}\ results, finding a difference of $\sim$0.1\,dex between the derived gravities (in which lower gravities
are found from the {\sc fastwind}\ analyzes). This warns us about the use of log\,{\em g}\,vs.\,log\,$T_{\rm eff}$\ to infer 
stellar ages and masses. However, as we are presenting a differential analysis based on results from the same code 
and techniques, our conclusions regarding the properties of O\,Vz stars should not be affected.

\subsection{Wind properties} \label{wind}

To investigate the wind properties of the analyzed sample, we present the derived
log\,{\em Q}\ values as a function of the effective temperature in Fig.~\ref{diagrama_q}.
From inspection of their location in this figure, one could conclude that the O\,Vz stars 
are concentrated at the lower envelope of the distribution (i.e., they have weaker winds). However, it is also interesting 
to have a closer look at the wind properties of both samples in the three temperature zones described in Sect.~\ref{tyg}.

It is particularly remarkable that, although a similar range in log\,{\em Q}\ is found in the three zones
(especially below 42000 K), the relative distribution of log\,{\em Q}\ values for both types of objects is
clearly dependent on the range of effective temperatures considered:

\begin{itemize}
 \item Zone 1: The Vz characteristic is not present in any stars, even in those
  for which low log\,{\em Q}\ values were obtained.
 \item Zone 2: Except for a few O\,V\ stars with values of log\,{\em Q}\ above $-$12.7, there 
 is no clear correlation in this region between wind strength and
 O\,Vz/O\,V\ classification. Intriguingly, there are a fair number of stars identified as O\,V\ with
 similar wind strengths (or upper limits in log\,{\em Q}) compared to some of the O\,Vz stars.
\item Zone 3: Although the O\,V\ stars generally have higher log\,{\em Q}\
  values than the O\,Vz stars, it is interesting that there are a few
  O\,Vz stars with relatively large log\,{\em Q}\ estimates (in terms of the
  range of values measured for the whole sample).
 \end{itemize}

Therefore, although wind strength seems to play a role in the
occurrence of the Vz classification (as expected), there seems to be
another factor involved, especially in the Zone~2 temperature range.

We would like to point out here that UV observations (in
the spectral range 1100\,--\,1800\,\AA{}) could help to better
constrain the wind properties of those stars for which
only upper limits have been obtained. This spectral range
contains some lines which are more sensitive to wind
variations than H$\alpha$ and \ion{He}{ii}\,$\lambda$4686, especially
in the weak wind regime.

\begin{table}[h!!!]
\caption{Number of O\,Vz and O\,V\ stars in different age ranges and temperature zones as estimated using the evolutionary tracks 
from \cite{brott} with an initial rotation rate of 171\,km\,s$^{-1}$\ (see Fig.~\ref{hrd}).}
\label{ages_table}
 \centering
\begin{tabular}{rcc@{\hspace{0.1cm}}|@{\hspace{0.1cm}}cc@{\hspace{0.1cm}}|@{\hspace{0.1cm}}cc@{\hspace{0.1cm}}|@{\hspace{0.1cm}}cc}
& \multicolumn{2}{c}{\textbf{\underline{ZONE 1}}}& \multicolumn{2}{c}{\textbf{\underline{ZONE 2}}}& \multicolumn{2}{c}{\textbf{\underline{ZONE 3}}}& \multicolumn{2}{c}{Total}\\ [1 ex]
Age & O\,Vz & O\,V & O\,Vz & O\,V & O\,Vz & O\,V&O\,Vz & O\,V\\
\hline 
& \multicolumn{2}{c|@{\hspace{0.1cm}}}{}& \multicolumn{2}{c|@{\hspace{0.1cm}}}{}& \multicolumn{2}{c|@{\hspace{0.1cm}}}{}& \multicolumn{2}{c}{}\\ [-2 ex]

$<$1\,Myr  & 0 & 0  & 4  & 1 & 7 & 0&11&1\\
1\,--\,2\,Myr     & 0 & 1  & 1  & 4 & 8 & 4&9&9\\
2\,--\,3\,Myr     & 0 & 2  & 7  & 3 & 7 & 3&14&8\\ 
3\,--\,4\,Myr     & 0 & 1  & 2  & 9 & 0 & 0&2&10\\ 
$>$\,4\,Myr       & 0 & 11 & 0  & 1 & 0 & 0&0&12\\
\hline
& \multicolumn{2}{c|@{\hspace{0.1cm}}}{}& \multicolumn{2}{c|@{\hspace{0.1cm}}}{}& \multicolumn{2}{c|@{\hspace{0.1cm}}}{}& \multicolumn{2}{c}{}\\ [-2 ex]
Total            & 0 & 15 & 14 & 18&22 & 7&36&40\\ 
\end{tabular}
\end{table}

\subsection{Luminosities and ages}\label{lya}

As the star evolves away from the ZAMS towards lower temperatures, its
gravity decreases and its stellar luminosity increases, so its wind
strength is also expected to increase. Since the intensity of the
\ion{He}{ii}\,$\lambda$4686 line depends strongly on this parameter
(see Sect.~\ref{introduction}), it is predicted that O stars evolving
from the ZAMS should lose their Vz characteristic.

A first view of the H--R diagram for the stars (Fig.~\ref{hrd})
indicates that the lower envelope of the distribution (with respect to
the ZAMS) is dominated by the O\,Vz stars, while the O\,V\ objects
concentrate in the upper envelope. Therefore, one could conclude that,
as a group, the O\,Vz stars in 30 Dor are younger and less luminous
(in an evolutionary sense) than the O\,V\ stars. However, a more
detailed inspection (see also Table~\ref{ages_table}) indicates the
presence of a non-negligible, unexpected number of O\,Vz objects at
relatively advanced ages (between 2 and 4\,Myr) and ten O\,V\ stars
similarly close to the ZAMS (below the 2\,Myr isochrone) as some of the
O\,Vz stars.

The latter feature can likely be explained by the rotational
velocities of the stars.  Most of the O\,V\ stars below the 2\,Myr
isochrone (computed from models with an {\em initial} $v$\,sin\,$i$\ of
171\,km\,s$^{-1}$) actually have $v$\,sin\,$i$\,$>$\,200\,km\,s$^{-1}$, and so they could be
somewhat more evolved objects. However, the large relative number of
O\,Vz to O\,V\ stars with inferred ages of between 2 and 4\,Myr is more
challenging to understand.

The three temperature zones described in Sect.~\ref{tyg} are also
indicated in Fig.~\ref{hrd}.  Interestingly, the O\,Vz and O\,V\ stars
in Zone~2 cover a similar range in luminosity, i.e., stars of both
types can be found having similar temperatures and luminosities.  The
O\,V\ stars in Zone~3 are mainly concentrated at higher luminosities,
but we also find O\,Vz stars in this luminosity regime; this seems to
indicate that other parameters/effects apart from age and luminosity
may play a role in the O\,Vz phenomenon.

When interpreting results from the H--R diagram, the effects of
visually-undetected binaries (or physically-unrelated, close
companions) on the measured photometry (hence the derived luminosity)
should be considered.  VFTS\,096 is an example of an O\,Vz star with
an erroneous luminosity (plotted as a red open square in
Fig.~\ref{hrd}). Inspection of the HST/WFC3 images revealed a nearby
companion of similar brightness within the 1.2\arcsec\ fibre aperture.
Due to their proximity, the two objects were not resolved in the
ground-based images used to obtain the photometry and, as a
consequence, the star appears to be overluminous when compared with
other objects with similar $T_{\rm eff}$\ and log\,{\em g}\ (e.g., VFTS\,601 and
746).

We have checked for similar cases among the O\,Vz stars located above
the 2\,Myr isochrone and found that VFTS\,096 is the only one with a
clear detection of nearby objects which may significantly influence
the measured photometry\footnote{VFTS\,110 and 398 also have close
  companions in the HST/WFC3 images. However, in both cases the {\em
    visual companion} is significantly fainter and so the expected effect on $m_{\rm
    v}$ is almost negligible.}. Even if we assume that VFTS\,096
only contributes half of the estimated luminosity, it would still
be located far away from the ZAMS (well above the 2\,Myr isochrone).
Therefore, although it cannot be completely ruled out, we consider it
very improbable that unresolved companions can explain the relatively
large number of O\,Vz stars with ages above 2\,Myr.  A more likely
explanation, related to the dependence of the wind strength with
metallicity, is presented in Sect.~\ref{test_fw}.

\subsection{Summary of results from the analysis}

In summary, the analysis of our sample of 38 O\,Vz and 46 O\,V\ stars
in 30\,Dor leads to the following main conclusions:
\begin{enumerate}
\item There do not appear to be significant differences in the derived
  gravities of the O\,Vz and O\,V\ stars. In particular, the O\,Vz
  stars do not concentrate at higher gravities relative to normal O
  dwarfs.
\item The O\,Vz stars tend to be closer to the ZAMS than O\,V\ stars,
  but we have also found a non-negligible number further away (with
  ages of 2-4\,Myr).
\item As a group, the O\,Vz stars tend to define the lower envelope of
  luminosities and wind strengths, but they are also found over the full
  range of values obtained for normal O\,V stars.
\item As expected, wind strength is an important parameter to be taken
  into account for the understanding of the Vz phenomenon.  In
  addition, the distinct temperature distribution found for the O\,Vz
  and O\,V\ samples indicates that $T_{\rm eff}$\ is the second key parameter.
  Also, the overlapping ranges of stellar parameters of both types of
  objects in the range 35000\,$\le$\,$T_{\rm eff}$\,$\le$\,40000\,K point
  towards one or more additional parameters playing a role.
\end{enumerate}

\section{The Vz phenomenon as predicted by {\sc fastwind} models}\label{test_fw}

In this section, we use synthetic spectra from the grid of {\sc fastwind}\ models to investigate 
the predicted effect of several spectroscopic parameters in the 
occurrence of the Vz characteristic. In particular, we concentrate on effective temperature, gravity, wind strength and
projected rotational velocity.

\subsection{Behaviour of the relevant He lines}\label{behave1}

\begin{figure*}[t!]
  \centering
   \includegraphics[scale=0.65, angle=90]{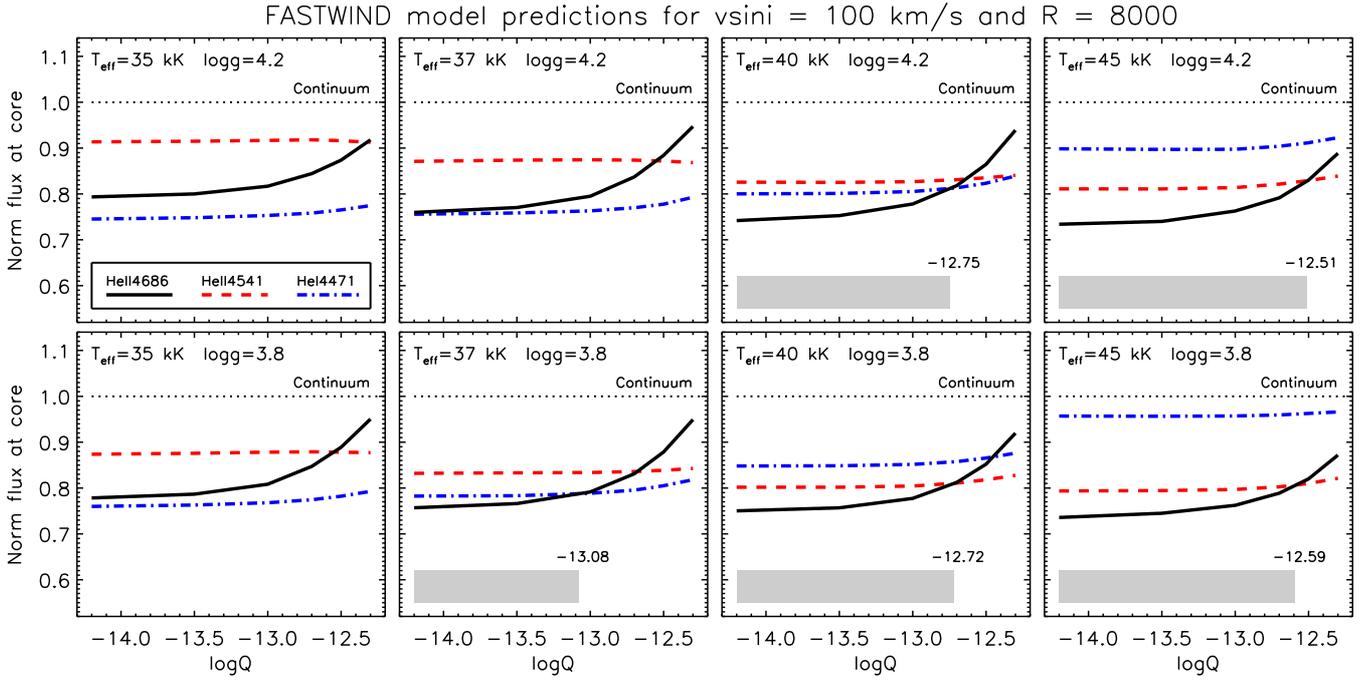}
  \caption{  \label{behav_lines} {\sc fastwind}\ predictions of the behavior of the lines involved 
  in the Vz phenomenon (\ion{He}{i}\,$\lambda$4471, \ion{He}{ii}\,$\lambda$4541 and \ion{He}{ii}\,$\lambda$4686) as a function of 
  the wind-strength Q-parameter for different ($T_{\rm eff}$,  log\,{\em g}) pairs. A fixed projected rotational velocity 
  ($v$\,sin\,$i$\,=\,100\,km\,s$^{-1}$) and resolution ($R$\,=\,8000) have been considered. The range of log\,{\em Q}\ values for which 
  a star would be classifed as a Vz star are indicated with the gray rectangle. The corresponding upper log\,{\em Q}\
  value leading to the Vz characteristic is also quoted.}
 \end{figure*}

An O dwarf is classified as Vz when \ion{He}{ii}\,$\lambda$4686 is
stronger in absorption than both \ion{He}{i}\,$\lambda$4471 and
\ion{He}{ii}\,$\lambda$4541. We therefore focussed the {\sc fastwind}\
predictions on these three lines.  Figure~\ref{behav_lines} shows the
variation of the normalized central intensities for the three lines as
a function of log\,{\em Q}, for representative ($T_{\rm eff}$, log\,{\em g}) pairs, which
span the range of our results in 30\,Dor (see Figs
\ref{gtd} and \ref{diagrama_q}).  In particular, we have considered
four $T_{\rm eff}$\ values and two log\,{\em g}\ values (organized in columns and
rows, respectively). The synthetic lines were degraded to the
resolving power of the VFTS Medusa spectra ($R$\,=\,8\,000), and
convolved with a rotational broadening profile corresponding to
$v$\,sin\,$i$\,=\,100\,km\,s$^{-1}$. We initially fixed $v$\,sin\,$i$\ to this value for
simplicity in the discussion; further investigation of the effect of
rotational broadening on the Vz characteristic is discussed in
Sect.~\ref{rotation_effect}.

 As expected, while there is a strong dependence of the depth of the
 \ion{He}{i}\,$\lambda$4471 and \ion{He}{ii}\,$\lambda$4541 lines with
 $T_{\rm eff}$\ (and to a lesser extent, log\,{\em g}), these lines are almost
 insensitive to variations in the wind strength (at the values
 corresponding to O dwarfs). In contrast, the flux at the core of
 \ion{He}{ii}\,$\lambda$4686 is only somewhat affected by changes in $T_{\rm eff}$\
 and log\,{\em g}, but varies strongly with log\,{\em Q}. These plots also allow us
 to illustrate why it is only possible to provide upper limits for the
 wind-strength parameter in a large fraction of the analyzed stars:
 below a certain value of log\,{\em Q}\ (between --13.0 and --13.5) the depth
 of \ion{He}{ii}\,$\lambda$4686 remains unaffected. A similar effect
 also occurs for H$\alpha$, the other (main) wind diagnostic line in
 the optical spectral range.

 As indicated above, the Vz characteristic is defined as
 \ion{He}{ii}\,$\lambda$4686 line being deeper than
 \ion{He}{ii}\,$\lambda$4541 and \ion{He}{i}\,$\lambda$4471. In
 Fig.~\ref{behav_lines} this translates to the black solid curve in a
 given panel being below the red dashed and blue dash-dotted lines.
 The range in log\,{\em Q}\ where this occurs for a given ($T_{\rm eff}$, log\,{\em g}) pair
 is indicated with a gray rectangle. The two main conclusions which
 can be extracted from inspection of these figures are:
\begin{itemize}
 \item Below a given $T_{\rm eff}$\ (which depends on the gravity) the Vz characteristic never occurs. \ion{He}{i}\,$\lambda$4471
 is always stronger in absorption than \ion{He}{ii}\,$\lambda$4686 due to the low temperature of the star, independent of 
 the wind strength. 
 \item At higher temperatures there is a certain log\,{\em Q}\ value below which the star would be
 classified as Vz. This upper limit in the wind-strength parameter seems to increase with temperature (i.e., 
 stars with higher temperatures require higher log\,{\em Q}\ values to lose their Vz characteristic), and
 also depends on log\,{\em g}\ for a given $T_{\rm eff}$.
\end{itemize}

The combined dependencies of the Vz characteristic on $T_{\rm eff}$, log\,{\em g}\
and log\,{\em Q}\ are shown in a more compact way in Fig.~\ref{limit_q_v100}.
The boundaries in log\,{\em Q}\ between the regions where O\,Vz and O\,V\
stars would be expected (below and above the lines, respectively) are
plotted as a function of $T_{\rm eff}$, for $v$\,sin\,$i$\,=\,100\,km\,s$^{-1}$\ and two
log\,{\em g}\ values.  The two curves vary similarly with $T_{\rm eff}$,
characterized by a strong dependence of log\,{\em Q}$_{\rm lim}$ at
intermediate temperatures, and an almost flat region above a certain
value (i.e., the boundary in log\,{\em Q}\ becomes almost insensitive to
temperature). In addition, the curve for log\,{\em g}\,=\,4.2 is shifted to
somewhat higher temperatures and log\,{\em Q}\ values compared to
log\,{\em g}\,=\,3.8.

Therefore, in addition to wind strength, the effective temperature and
gravity of the star are two important parameters to be taken into
account to interpret the Vz phenomenon from an evolutionary point of
view. This may be especially important in the range of temperatures
where there is a strong dependence of $Q_{\rm lim}$ with $T_{\rm eff}$\ and
log\,{\em g}. For example, a star with $T_{\rm eff}$\,=\,38000 K (log\,$T_{\rm eff}$=\,4.58),
$v$\,sin\,$i$\,=\,100\,km\,s$^{-1}$\ and log\,{\em Q}\,=\,--12.8 (an intermediate value for
the 30\,Dor sample, see Fig.~\ref{diagrama_q}) would be classified as
V if log\,{\em g}\,=\,4.2, and as Vz if log\,{\em g}\,=\,3.8.  Note that the star
with log\,{\em g}\,=\,4.2\,dex would still be classified as O\,V, even if the
wind strength is much lower. While this example contradicts what might
be expected from an evolutionary point of view, it can be explained by
taking into account the relative dependencies of the three diagnostic
lines with $T_{\rm eff}$, log\,{\em g}, and log\,{\em Q}.

In Sect.~\ref{obs_vs_fw} we discuss the implications of the {\sc fastwind}\
predictions on the interpretation of our results for the samples in
30\,Dor in more detail. However, we first investigate a fourth
important parameter, the rotational broadening ($v$\,sin\,$i$).

\subsection{The role of rotational broadening on the Vz phenomenon}\label{rotation_effect}

Rotational broadening is expected to influence the depths of the
\ion{He}{i} and \ion{He}{ii} lines differently \citep[e.g.,]
[]{markova11}. As the Vz classification is based on the relative
depths of \ion{He}{i}\,$\lambda$4471, \ion{He}{ii}\,$\lambda$4541,
and \ion{He}{ii}\,$\lambda$4686, we also investigated the influence
that rotation may have on its occurrence. 

\begin{figure}[h!]
  \centering
   \includegraphics[scale=0.50, angle=0]{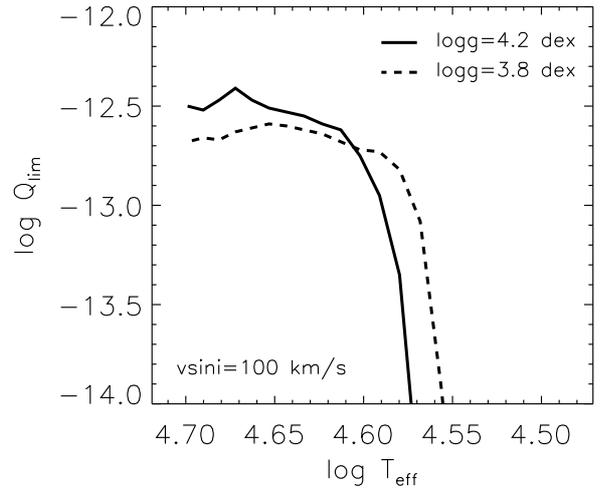}
   \caption{ \label{limit_q_v100} Upper limits in the Q-parameter, as
     a function of $T_{\rm eff}$, below which Vz objects are expected (for
     $v$\,sin\,$i$\,=\,100\,km\,s$^{-1}$\ and two different values of log\,{\em g}).}
 \end{figure}

\begin{figure}[h!]
  \centering
   \includegraphics[scale=0.50, angle=0]{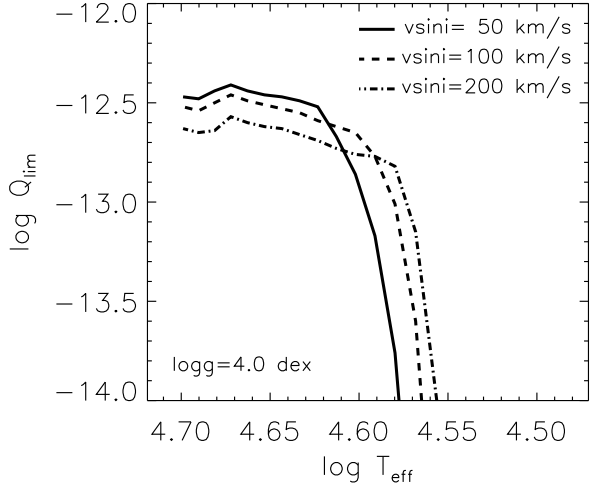}
  \caption{\label{limit_q} Same as Fig.~\ref{limit_q_v100} but for log\,{\em g}\,=\,4.0\,dex and three different
  values of $v$\,sin\,$i$.}
 \end{figure}

In Fig.~\ref{limit_q} we plot a similar comparison to that shown in
Fig.~\ref{limit_q_v100}, but now for a fixed gravity and three
different $v$\,sin\,$i$\ values.  The curves in the figure confirm an
important influence of rotational broadening, which is comparable in
impact to that of the gravity (although with different causes).  In particular,
the region in which there is a strong dependence of the $Q_{\lim}$ boundary
with $T_{\rm eff}$, moves to lower $T_{\rm eff}$\ with increasing $v$\,sin\,$i$.

We conclude that higher values of $v$\,sin\,$i$\ generally favor the
presence of O\,Vz stars at relatively low temperatures (below
$T_{\rm eff}$\,$\sim$\,38000 K), with the opposite effect in the high $T_{\rm eff}$\
regime (where a lower log\,{\em Q}\ value is required to lose the Vz
characteristic).

This result can be easily understood taking into account the different
(relative) effect that rotational broadening produces on the three
diagnostic lines. In particular, one should remember that intrinsic
Stark broadening in the \ion{He}{i} line is quadratic, while it is
linear in the \ion{He}{ii} lines (and that the intrinsic width of
\ion{He}{ii}\,$\lambda$4541 is slightly broader than
\ion{He}{ii}\,$\lambda$4686).  The narrower the line, the more it is
affected by rotational broadening. This explains why, at relatively
low temperatures (when \ion{He}{i}\,$\lambda$4471 is stronger in
absorption than the other two \ion{He}{ii} lines), rotation favors
the Vz characteristic for a given value of log\,{\em Q}. At higher
temperatures, when the important diagnostic lines in the definition of
the Vz phenomenon are \ion{He}{ii}\,$\lambda$4541 and
\ion{He}{ii}\,$\lambda$4686, the effect works in the opposite
direction.

\subsection{Summary of {\sc fastwind} predictions}\label{summary_fw}

To sum up, the main conclusions from the {\sc fastwind}\ model predictions
concerning the Vz characteristic are:

\begin{enumerate}
\item Wind strength and effective temperature are the key parameters
  in understanding the Vz phenomenon. However, other parameters such
  as gravity and projected rotational velocity are also important in
  defining the occurrence of this characteristic for a given wind
  strength and temperature.
\item At low temperatures ($T_{\rm eff}$\,$\lesssim$\,35000\,K) only O\,V\
  stars are expected, independent of the wind strength. The
  temperature is not sufficiently high for \ion{He}{ii}\,$\lambda$4686
  to be stronger in absorption than \ion{He}{i}\,$\lambda$4471, even
  for very low values of log\,{\em Q}.
\item At intermediate temperatures
  (35000\,$\lesssim$\,$T_{\rm eff}$\,$\lesssim$\,40000\,K) occurrence of the
  Vz characteristic for a given value of log\,{\em Q}\ strongly depends on
  the specific values of $T_{\rm eff}$, log\,{\em g}\ and $v$\,sin\,$i$. In this region,
  {\sc fastwind}\ models predict that high gravities and low $v$\,sin\,$i$\ values
  favor the presence of normal O\,V\ stars, even for objects with
  relatively low values of log\,{\em Q}. The relative fraction of Vz to V stars
  in this region is expected to increase towards the upper limit in
  $T_{\rm eff}$.
\item At high temperatures ($T_{\rm eff}$\,$\gtrsim$\,40000\,K) the boundary
  in log\,{\em Q}\ between O\,Vz and O\,V\ stars becomes almost insensitive to
  $T_{\rm eff}$, and relatively high values of log\,{\em Q}\ are needed to make a
  star of a given $T_{\rm eff}$\ lose its Vz characteristic. In this region,
  high gravities and low $v$\,sin\,$i$\ values favor the presence of Vz
  stars.
\end{enumerate}

\section{The Vz phenomenon: observations vs. model atmosphere predictions}\label{obs_vs_fw}

Our investigation of the O\,Vz phenomenon as predicted by the
{\sc fastwind}\ models allows us to further interpret the results from our
analysis of the O\,Vz and O\,V\ samples in 30\,Dor.

  \begin{figure*}[ht!!!!!!]
  \centering
   \includegraphics[scale=0.60, angle=90]{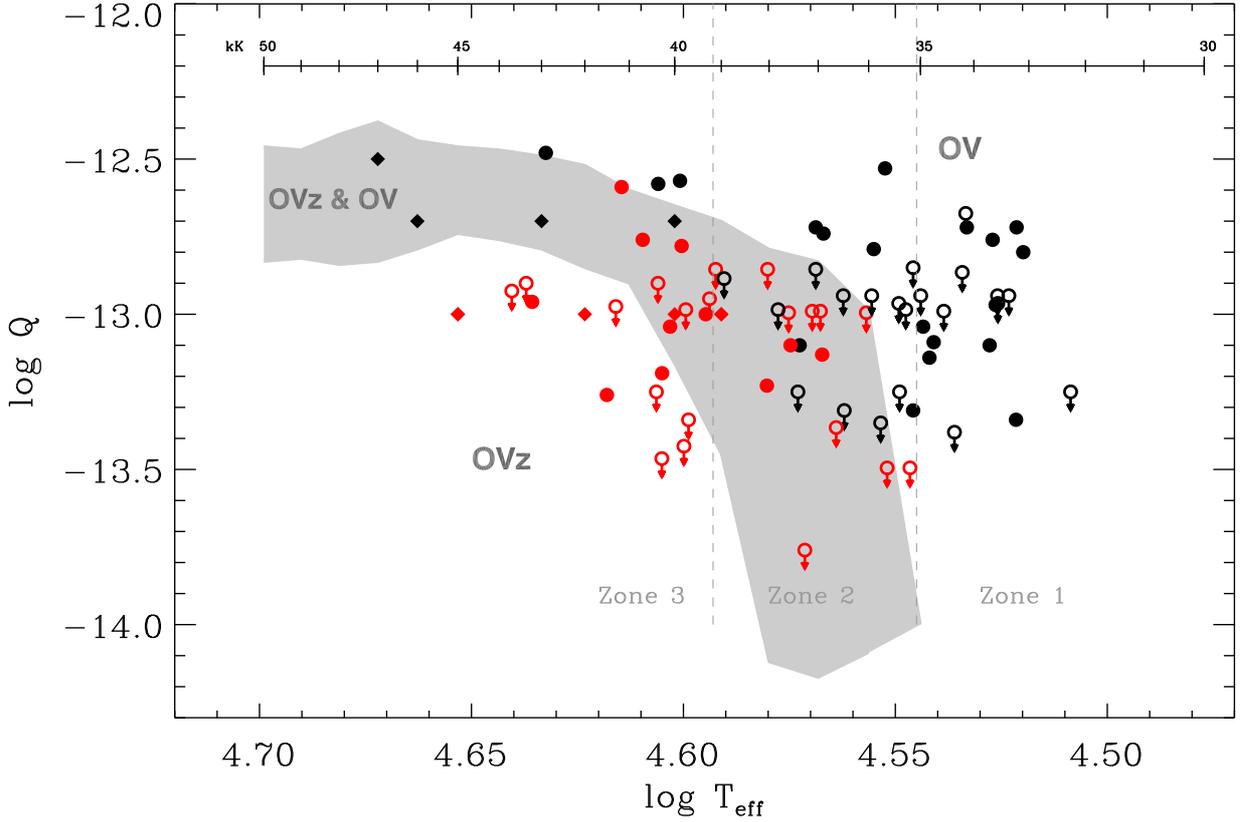}
   \caption{ \label{low_rot} {\sc fastwind}\ predictions for Vz behavior
     with log\,{\em Q}\ as a function of $T_{\rm eff}$, for models with
     3.8\,$\le$\,log\,{\em g}\,$\le$\,4.2\,dex, and
     50\,$\le$\,$v$\,sin\,$i$\,$\le$\,300\,km\,s$^{-1}$, compared to the results from
     our analyzes with the IACOB-GBAT.  The shaded area indicates the
     region where we expect both O\,Vz and O\,V\ stars from our
     analysis (see text), whereas we expect only O\,V/O\,Vz stars
     above/below this area.  Symbols as used in Fig.~\ref{diagrama_q}.}
\end{figure*}

In the previous section we have shown that the log\,{\em Q}\,vs.\,log\,$T_{\rm eff}$\
diagram provides a useful way to illustrate the predicted dependence
of the occurrence of the Vz phenomenon on $T_{\rm eff}$, log\,{\em g}, log\,{\em Q}\ and
$v$\,sin\,$i$. In Fig.~\ref{low_rot} we combine our results for the O\,Vz and
O\,V\ stars in 30\,Dor, with the information extracted from the
{\sc fastwind}\ model predictions. This figure is similar to
Fig.~\ref{diagrama_q}, but now includes the gray shaded area, which
indicates the region where both Vz and V stars are expected (depending
on the specific values of log\,{\em g}\ and $v$\,sin\,$i$). The boundaries of the
shaded area were obtained from the $Q_{\rm lim}$ values contained in
curves such as those in Figs.~\ref{limit_q_v100} and \ref{limit_q}
(for 3.8\,$\le$\,log\,{\em g}\,$\le$\,4.2 and
50\,$\le$\,$v$\,sin\,$i$\,$\le$\,300\,km\,s$^{-1}$). Below/above this area, only O\,Vz/O\,V\ stars are expected.

The first thing to remark on in Fig.~\ref{low_rot} is the excellent
agreement between the V or Vz classification of the analyzed stars and
the {\sc fastwind}\ predictions. Clean samples of O\,V and O\,Vz
stars are found above and below the shaded region (respectively), but
there is a mixture of both types of objects within that region.

With this figure in mind (as well as Figs.~\ref{limit_q_v100} and
\ref{limit_q}) we can revisit the distribution of O\,Vz and O\,V\ stars
in the log\,{\em g}\,vs.\,log\,$T_{\rm eff}$\ and H--R diagrams (Figs.~\ref{gtd} and
\ref{hrd}, respectively). It is now clear why no O\,Vz stars are found
in Zone~1, while the dominance of O\,Vz stars in Zone~3 can be
understood by taking into account that most of the O dwarfs in 30\,Dor
do not have a strong enough wind to break the Vz characteristic
(likely a consequence of the low metallicity of the LMC). Finally, the
strong dependence of the Vz characteristic on the specific combination
of $T_{\rm eff}$, log\,{\em g}, log\,{\em Q}, and $v$\,sin\,$i$\ for stars with
35000\,$\lesssim$\,$T_{\rm eff}$\,$\lesssim$\,40000\,K, explains the mixture
of O\,Vz and O\,V stars found in Zone~2.

It is also interesting to realize that the unexpected distribution of
gravities for the objects in Zone~2 (e.g., Fig.~\ref{gtd}) may be
explained by taking into account the {\sc fastwind}\ predictions. In
Sect.~\ref{tyg} it was noted that O\,Vz stars in this zone tend to
concentrate at lower gravities than normal O dwarfs (see
Fig.~\ref{gtd}). This result is, in some sense, expected from the fact
that {\sc fastwind}\ models predict that high gravities favor the presence
of O\,V\ stars in this range of temperatures, even for relatively low
values of log\,{\em Q}\ (see Fig.~\ref{limit_q_v100}).

We therefore conclude that the global distribution of O\,Vz and O\,V\
stars in 30\,Dor in the log\,{\em g}\,vs.\,log\,$T_{\rm eff}$,
log\,{\em Q}\,vs.\,log\,$T_{\rm eff}$, and H--R (log\,$L$\,vs.\,log\,$T_{\rm eff}$) diagrams
(and even most of the unexpected cases) can be explained as a natural
combination of stellar parameters. Similar arguments will likely be
applicable to the variety of results for O\,Vz stars in the literature
(see notes in Sect.~\ref{introduction}).

\section{Summary and conclusions}\label{conclusion}

The O\,Vz stars, a subclass of the O-type dwarfs characterized by
having \ion{He}{ii}\,$\lambda$4686 stronger in absorption than any
other He line in their blue-violet spectra, have been suggested to be
stars on or near the ZAMS \citep[][]{w09}. In particular, there are
empirical arguments that O\,Vz stars are younger and subluminous, with
higher gravities and weaker winds, when compared to normal O dwarfs.
These properties would make O\,Vz stars of considerable interest to
advance our knowledge of the physical properties of massive stars in
the very first stages of their lives, an active research field which
is still open to debate.

The VFTS has provided a unique opportunity to investigate the proposed
hypotheses about the nature of the O\,Vz stars in more detail.  This
is because of the large number of O\,Vz stars discovered in 30\,Dor,
combined with the excellent quality (in terms of resolving power, S/N
ratio, and spectral coverage), homogeneity, and multi-epoch nature of
the VFTS data. This has enabled us to perform a comprehensive
quantitative analysis of a statistically-meaningful sample of (very
likely single) O\,Vz and O\,V\ stars in the same star-forming region
for the first time.

The ultimate goal of this work was to investigate the postulated
different (younger) evolutionary stage of O\,Vz stars in comparison
with normal O dwarfs. To address this, we obtained the stellar
and wind parameters of a sample of 38 O\,Vz stars (plus a control
sample of 46 O\,V\ stars) in 30\,Dor by means of standard techniques,
using the {\sc fastwind}\ stellar atmosphere code and the IACOB-GBAT, a
grid-based tool developed for automated quantitative analysis of
optical spectra of O stars. The derived parameters have been located
in three diagrams of interest, namely the log\,{\em g}\,vs.\,log\,$T_{\rm eff}$,
log\,{\em Q}\,vs.\,log\,$T_{\rm eff}$\ and H--R diagrams, which have been used to
investigate if the O\,Vz stars have lower luminosities, higher gravities,
and weaker winds than the O\,V\ stars, and if they are also closer to
the ZAMS (i.e., younger), in the frame of a differential analysis.

From inspection of these diagrams we conclude that O\,Vz stars do not
have higher gravities, but they define the lower envelope of the
distribution of stars in the H--R and log\,{\em Q}\,vs.\,log\,$T_{\rm eff}$\ diagrams
(i.e., are generally younger and have weaker winds\footnote{We
  note that, given our limitations to determine the wind strength
  below a certain threshold, this does not confirm nor reject a
  possible relation between some O\,Vz stars and the weak-wind stars}
than normal O dwarfs). However, we have also found some results that
seem to challenge the postulated nature of the O\,Vz stars: (a) not
all O\,Vz stars in our sample are less luminous and younger than O\,V\
stars with the same spectral types, (b) there is a non-negligible
number of O\,Vz stars with relatively advanced ages (of 2-4\,Myr), and
(c) at intermediate temperatures (between approx. 35000 and 40000\,K)
there is not a clear correlation between wind strength and Vz (or V) nature.
We note that these findings agree with the diversity of results found
by other authors in O\,Vz stars within the Galaxy and the SMC (see
Sect.~\ref{introduction}).

Interestingly, the two samples have distinct distributions in
temperature/spectral type. The O\,Vz stars concentrate at intermediate
and high temperatures, while lower temperatures are dominated by the O\,V\
stars. In addition, we have found that the relative distribution of
gravities, luminosities, and wind-strength in both types of objects
depends on the effective temperature considered. This result indicates
that other parameters, apart from wind strength and effective
temperature, play a secondary role in the occurrence of the Vz
characteristic (such as gravity, luminosity and rotational velocity).
\textit{As a consequence of the interpretation of the Vz phenomenon in
  terms of proximity to the ZAMS (i.e., age), it is more complex than
  initially expected.} We should also keep in mind that because
rotational broadening has important effects on the way that massive
stars evolve away from the ZAMS, two O dwarfs located at the same
place in the H--R diagram could actually correspond to two different
evolutionary phases.

These findings are confirmed by {\sc fastwind}\ predictions of the relative
behavior (as a function of $T_{\rm eff}$, log\,{\em g}, log\,{\em Q}, and $v$\,sin\,$i$) of the
three main diagnostic lines which define the Vz subclass. This has
allowed us to identify the importance of considering the specific
combination of these four parameters (in addition to the location of
the star in the H--R diagram) for a more complete understanding of the
nature of the O\,Vz phenomenon from an evolutionary point of view.
This is especially critical for stars with effective temperatures
between 35000 and 40000\,K. In this temperature range, a star with a
modest (or even low) wind strength will appear as O\,Vz or O\,V\
depending on the specific values of $T_{\rm eff}$, log\,{\em g}, and $v$\,sin\,$i$. In
particular, we have shown that lower gravities and higher projected
rotational velocities favor the occurrence of the Vz characteristic
for a given value of log\,{\em Q}\ at intermediate temperatures.

Other predictions of interest are: (a) below $\sim$35000\,K, \ion{He}{i}\,$\lambda$4471 always dominates 
in absorption over \ion{He}{ii}\,$\lambda$4686 due to the low temperature of the star; as a consequence, 
no O\,Vz stars are expected to be found at these temperatures, even if the star has a weak wind; 
(b) above $\sim$40000\,K, the boundary in log\,{\em Q}\ between O\,Vz and O\,V\ stars becomes 
almost insensitive to $T_{\rm eff}$, and relatively high values of log\,{\em Q}\ are needed to make a star of 
a given $T_{\rm eff}$\ lose its Vz characteristic. In this region, high gravities and low $v$\,sin\,$i$\ values 
favor the presence of Vz stars. 

The comparison of results from the spectroscopic analysis and the
{\sc fastwind}\ predictions has shown that the distribution of our O\,Vz
and O\,V\ stars in the log\,{\em g}\,vs.\,log\,$T_{\rm eff}$, log\,{\em Q}\,vs.\,log\,$T_{\rm eff}$,
and H--R diagrams can be naturally explained as a combination of
stellar parameters. However, one should also keep in mind that, as
indicated by \cite{walborn13}, the high binary frequency of
the O stars might provide an alternative origin for some O\,Vz
spectra: composite spectra of (morphologically normal) early and late
O-type dwarfs may produce an apparent mid O\,Vz morphology.

We propose an explanation to the surprisingly large number of O\,Vz
stars found in 30\,Dor, and why some of them are found an unexpectedly
large distance away from the ZAMS (above the 2\,Myr isochrone).  Our
hypothesis refers to metallicity: \textit{the low metal content of the
  region implies that most of the O-dwarfs in 30\,Dor do not have a
  wind strong enough to break the Vz characteristic, even in slightly
  evolved stars.}  Following from this, a lower percentage of Vz stars
might be expected to be found away from the ZAMS in the Galaxy (due to
the greater metallicity).  In any case, while it seems that some are
not ZAMS objects, the O\,Vz stars remain important objects to be
considered in the investigation of the physical properties of the very
early phases of massive stars in the Universe.

\begin{acknowledgements}
CS-SJ, SS-D and AH acknowledge financial support
from the Spanish Ministry of Economy and Competitiveness (MINECO) under
the grants AYA2010-21697-C05-04, Consolider-Ingenio 2010 CSD2006-00070,
and Severo Ochoa SEV-2011-0187, and by the Canary Islands Government under grant PID2010119.
FN and MG acknowledge support by the Spanish MINECO under grants
AYA2010-21697-C05-01 and FIS2012-39162-C06-01.
STScI is operated by the Association of Universities for Research in Astronomy, Inc., under NASA contract NAS 5-26555.
JMA acknowledges support from the Spanish Government Ministerio de Educaci\'on y Ciencia through grants AYA2010-15081 
and AYA2010-17631 and the Consejer{\'\i}a de Educaci\'on of the Junta de Andaluc{\'\i}a through grant P08-TIC-4075.
The publication is supported by the Austrian Science Fund (FWF). 
\end{acknowledgements}


 \bibliographystyle{aa}
\bibliography{cssj_ref}



\appendix
\section{Tables}\label{apendB}
Tables \ref{tab_vz} and \ref{tab_v} present the stellar parameters
from our HHe analysis of the O\,Vz and O\,V\ samples.
Table~\ref{tab_teff} presents first estimates for the stars where it
was also necessary to use nitrogen lines in the analysis due to weak
\ion{He}{i} lines and/or strong nebular contamination.
\begin{landscape}
\begin{table}[t]
\centering
\caption{\label{tab_vz} Stellar and wind parameters obtained from quantitative analysis 
of our sample of O\,Vz stars.} 
\begin{tabular}{crlccccccccccccl}
\hline\hline
\multicolumn{15}{c}{} \\ [-2 ex]
VFTS&SpT& LC&$M_v^{(1)}$&$v$\,sin\,$i$&$T_{\rm eff}$&$\Delta$$T_{\rm eff}$&log\,{\em g}&$\Delta$log\,{\em g}$^{(2)}$& log\,{\em g}$_c^{(3)}$&Y(He)$^{(4)}$&log\,{\em Q}&$\Delta$log\,{\em Q}&$\log{L/L_{\odot}}^{(5)}$ &Comments$^{(6)}$\\
 && &&[$\rm{kms^{-1}}$]&[K]&[K]&[dex]&[dex]&[dex]& &[dex]&[dex] &[dex]& \\
\hline
\multicolumn{15}{c}{} \\ [-1.5 ex]

355&O4&V((n))((fc))z&-5.1&135&43400&  600&3.84&0.10&3.85&0.09& $<$-12.9&  --&5.52& NC\\
418&O5&V((n))((fc))z&-4.4&135&43200& 1700&4.09&0.13&4.10&0.09& -13.0&  0.3&5.24& SB?\\
511&O5&V((n))((fc))z&-4.9&105&43700& 1700&4.25&0.11&4.25&0.10&$<$-12.9& --&5.46& SB1s\\
398&O5.5&V((n))((f))z&-5.1& 65&41200&  1000&4.03&0.10&4.03&$<$0.06& -12.6&  0.2&5.47& SBvs\\
601&O5-6&V((n))z&-5.4&125&40300&  500&3.93&0.10&3.94&0.09& -13.2&  0.4&5.55& ...\\
096&O6&V((n))((fc))z&-5.7&125&40100&  500&3.90&0.10&3.91&0.09& -13.0&  0.2&5.67& SBvs VM2\\
110&O6&V((n))z&-5.0&150&39900& 1000&3.86&0.10&3.88&$<$0.06& -12.8&  0.2&5.40& VM2?\\
117&O6:&Vz&-4.1& 75&41300& 1500&4.14&0.16&4.14&0.12&$<$-13.0& --&5.02& SB?\\
356&O6:&V(n)z&-4.5&215&39300& 1300&3.99&0.13&4.03&0.10&$<$-13.0& --&5.14& SB?\\
470&O6:&V((f))z&-4.1& 75&39300&  600&3.93&0.10&3.94&0.10& -13.0&  0.5&4.97& ...\\
472&O6&Vz&-4.1& 40&40400&  900&4.12&0.12&4.12&0.11&$<$-12.9& --&5.01& ...\\
488&O6&V((f))z&-4.8& 55&40700&  700&3.87&0.10&3.87&0.09& -12.8&  0.2&5.33& ...\\
536&O6&Vz&-4.4& 40&41500& 1500&4.23&0.15&4.23&0.09& -13.3&  0.6&5.19& SB?\\
089&O6.5&V((f))z Nstr&-4.3& 50&39700&  700&4.02&0.12&4.02&0.13&$<$-13.3& --&5.09& ...\\
123&O6.5&Vz&-4.0& 65&40400&  700&4.10&0.12&4.10&0.13&$<$-13.3& --&4.99& SB?\\
549&O6.5&Vz&-4.3&110&39800& 1200&4.04&0.16&4.05&0.09&$<$-13.0& --&5.09& SB?\\
761&O6.5&V((n))((f))z Nstr&-4.1&110&40300&  700&4.15&0.10&4.16&0.18&$<$-13.5& --&4.99& NC\\
380&O6-7&Vz&-3.9& 65&39100&  700&4.13&0.10&4.13&$<$0.09&$<$-12.9& --&4.92& ...\\
392&O6-7&V((f))z&-4.5& 40&37600&  800&3.87&0.10&3.87&0.10& -13.1&  0.2&5.11& ...\\
706&O6-7&Vnnz&-4.2&375&38000& 1200&3.80&0.13&3.95&0.11&$<$-12.9& --&5.02& ...\\
722&O7&Vnnz&-4.1&405&36600&  800&3.84&0.10&4.01&0.11&$<$-13.4& --&4.91& SB? NC\\
724&O7&Vnnz&-4.3&370&37600& 3300&3.78&0.41&3.93&0.19&$<$-13.0& --&5.01& NC \\
849&O7&Vz&-4.1& 95&39800&  600&4.16&0.11&4.17&0.11&$<$-13.4& --&5.02& NC\\
751&O7-8&Vnnz&-4.4&360&36000& 1500&3.90&0.25&4.01&0.10&$<$-13.0& --&5.01& NC\\
266&O8&V((f))z&-4.3& 40&38000&  500&4.01&0.10&4.01&0.10& -13.2&  0.3&5.05& ...\\
014&O8.5&Vz&--& 90&37100&  600&3.91&0.10&--&$<$0.06&$<$-13.0& --&--& SBs\\
168&O8.5&Vz&-4.1& 40&37300&  500&4.02&0.10&4.02&0.10&$<$-13.8& --&4.92& SB?\\
252&O8.5&Vz&-3.6&100&37000&  500&4.21&0.10&4.22&0.11&$<$-13.0& --&4.73& ...\\
638&O8.5&Vz&-3.5& 45&36900&  500&4.20&0.10&4.20&0.10& -13.1&  0.3&4.68& ...\\
067&O9.5&Vz&-3.3& 40&35000& 1100&4.12&0.19&4.12&0.08&$<$-13.5& --&4.56& SB?\\
132&O9.5&Vz&-3.6& 40&35600&  700&4.18&0.10&4.18&0.10&$<$-13.5& --&4.71& ...\\

\hline
\multicolumn{5}{l}{} \\ [-1.5 ex]
\multicolumn{15}{l}{$^{(1)}$ Values taken from Ma\'iz Apell\'aniz et al. (in prep); the estimated average uncertainties are $\rm{\Delta M_V}$\,=\,0.3, accounting for uncertainties in the distance modulus \citep[see][]{gibson}}\\
\multicolumn{15}{l}{and the apparent magnitudes.}\\
\multicolumn{15}{l}{$^{(2)}$Formal errors adopting a minimum value of 0.1\,dex. } \\
\multicolumn{15}{l}{$^{(3)}\, \rm{\log{g_c}=\log{[g+(v\sin{i})^2/R_*]}}$ \citep[see][]{h92,rph04} }\\
\multicolumn{5}{l}{$^{(4)}$ $\rm{\Delta}$Y(He)\,=\,0.02}\\
\multicolumn{15}{l}{$^{(5)}$ The combined uncertainties in $M_V$ and $T_{\rm eff}$\ (from the IACOB-GBAT analysis) lead to an
estimated average error in luminosity of $\Delta \log{L/L_{\odot}}$\,=\,0.13\,dex.}\\
\multicolumn{15}{l}{$^{(6)}$ Relevant comments from Table~1 of \cite{walborn13}: SB, spectroscopic binary;
1, single lined; 2, double lined; s, small amplitude (10\,--\,20\,km\,s$^{-1}$ ); vs, very small amplitude}\\
\multicolumn{15}{l}{($<$10\,km\,s$^{-1}$ ); SB?, stellar absorption displaced from nebular emission lines but no radial-velocity variation measured; SB2?, confirmed SB with possible second companion or unconfirmed}\\
\multicolumn{15}{l}{SB but two companions visible in the line-of-sight; VMn, visual multiple of n components within the 1.2\arcsec\ Medusa fibre, as determined from HST/WFC3 images; NC, no WFC3 coverage.}\\
\end{tabular}
\end{table}
\end{landscape}
\begin{landscape}
\begin{table}[t]
\caption{\label{tab_v} Stellar and wind parameters obtained from quantitative analysis 
of our sample of O\,V\ stars. }
\centering
\begin{tabular}{crlcccccccccccc}
\hline\hline
\multicolumn{15}{c}{} \\ [-2 ex]

VFTS&SpT& LC&$M_v^{(1)}$&$v$\,sin\,$i$&$T_{\rm eff}$&$\Delta$$T_{\rm eff}$&log\,{\em g}&$\Delta$log\,{\em g}$^{(2)}$& log\,{\em g}$_c^{(3)}$&Y(He)$^{(4)}$&log\,{\em Q}&$\Delta$log\,{\em Q}&$\log{L/L_{\odot}}^{(5)}$ &Comments$^{(6)}$\\
 && &&[$\rm{kms^{-1}}$]&[K]&[K]&[dex]&[dex]&[dex]& &[dex]&[dex] &[dex] &\\
\hline 
\multicolumn{15}{c}{} \\ [-1.5 ex]

385&O4-5&V((n))((fc))&-5.2&120&42900& 1700&3.86&0.10&3.87&0.09& -12.5&  0.2&5.55& SBs\\
491&O6&V((fc))&-5.1& 50&40400&  800&3.84&0.08&3.84&0.10& -12.6&  0.2&5.43& SB?\\
746&O6&Vnn&-4.8&275&39900& 1200&3.86&0.10&3.92&0.08& -12.6&  0.2&5.29& ...\\
484&O6-7&V((n))&-5.4&120&35700&  700&3.67&0.10&3.68&$<$0.07& -12.5&  0.2&5.41& ...\\
770&O7&Vnn&-4.2&350&37800& 1100&3.95&0.15&4.06&0.10&$<$-13.0& --&4.98& ...\\
285&O7.5&Vnnn&-3.9&600&35300&  900&3.63&0.10&4.08&0.14&$<$-13.0& --&4.77& ...\\
065&O8&V(n)&-3.8&165&37100& 1100&4.06&0.16&4.08&0.10&$<$-12.9& --&4.80& ...\\
249&O8&Vn&-3.8&300&36500&  800&4.04&0.11&4.11&0.10&$<$-13.3& --&4.78& ...\\
494&O8&V(n)&-4.2&230&38900& 1700&4.20&0.23&4.21&0.09&$<$-12.9& --&5.03& SB2?\\
611&O8&V(n)&-3.8&210&37400&  900&4.09&0.14&4.13&0.10&$<$-13.3& --&4.79& ...\\
768&O8&Vn&-4.7&290&35100& 1200&3.88&0.18&3.95&0.10& -13.3&  0.6&5.09& SB2? NC\\
130&O8.5&V((n))&-4.5&170&36500& 1330&4.09&0.19&4.11&0.08&$<$-12.9& --&5.06& ...\\
154&O8.5&V&-5.0& 55&37400&  700&4.12&0.13&4.12&0.09& -13.1&  0.3&5.30& SBs\\
361&O8.5&V&-5.0& 70&36900&  700&4.07&0.10&4.07&$<$0.06& -12.7&  0.1&5.27& ...\\
597&O8-9&V(n)&-4.1&210&35400&  700&3.90&0.11&3.94&0.11&$<$-13.3& --&4.87& ...\\
074&O9&Vn&-3.7&265&35100& 1300&4.18&0.21&4.23&0.10&$<$-12.9& --&4.69& ...\\
138&O9&Vn&-3.5&350&34600&  900&4.10&0.14&4.20&0.10&$<$-13.0& --&4.60& SB2?\\
280&O9&V((n))&-4.2&150&34400&  600&3.82&0.10&3.85&0.10&$<$-13.4& --&4.88& ...\\
419&O9:&V(n)&-4.8&145&33100&  900&3.61&0.10&3.64&0.14& -12.8&  0.2&5.07& ...\\
483&O9&V&--& 40&33700&  900&4.09&0.11&--&0.11& -12.8&  0.2&--& SB?\\
493&O9&V&-4.4&200&37100&  1000&4.25&0.10&4.27&$<$0.08& -12.7&  0.2&5.06& ...\\
521&O9&V(n)&-5.0&150&34800&  600&4.12&0.10&4.13&0.09& -13.1&  0.2&5.21& VM2\\
892&O9&V&-4.0& 40&35800&  600&3.98&0.10&3.98&0.11&$<$-13.4& --&4.85& NC\\
250&O9.2&V((n))&-3.9&155&35400&  805&4.12&0.15&4.14&0.09&$<$-13.0& --&4.76& SB?\\
704&O9.2&V(n)&--&240&34200& 1500&3.98&0.22&--&0.09&$<$-12.7& --&--& SB?\\
775&O9.2&V&-3.7& 40&35900& 1300&4.14&0.20&4.14&0.12&$<$-12.9& --&4.72& SB? NC\\
149&O9.5&V&-3.7&125&35000& 1400&4.12&0.24&4.13&0.12&$<$-12.9& --&4.68& ...\\
498&O9.5&V&-4.3& 40&33200&  800&4.12&0.15&4.12&0.09& -13.3&  0.5&4.88& ...\\
560&O9.5&V&-3.3& 40&33600& 1200&4.20&0.16&4.20&0.09&$<$-12.9& --&4.52& ...\\
582&O9.5&V((n))&--&115&35000&  800&4.27&0.10&--&$<$0.08& -13.0&  0.3&--& ...\\
592&O9.5&Vn&-3.7&295&33600& 1000&4.23&0.13&4.28&$<$0.10&$<$-13.0& --&4.69& ...\\
649&O9.5&V&-3.7&105&34800&  600&4.18&0.10&4.19&0.09& -13.1&  0.2&4.71& ...\\
660&O9.5&Vnn&-4.0&515&32300& 1000&3.95&0.16&4.15&0.11&$<$-13.3& --&4.73& ...\\
677&O9.5&V&-4.4& 40&35900& 1100&4.20&0.13&4.20&$<$0.07& -12.8&  0.2&5.03& VM3\\
679&O9.5&V&-3.9& 40&33200&  900&4.10&0.15&4.10&$<$0.07& -12.7&  0.2&4.72& SB?\\
778&O9.5&V&-4.0&125&34200& 1400&4.20&0.22&4.19&0.09&$<$-12.9& --&4.81& ...\\
369&O9.7&V&-3.7& 40&33400& 1200&4.10&0.18&4.10&$<$0.09&$<$-12.9& --&4.67& ...\\
554&O9.7&V&--& 45&34100&  800&4.26&0.10&--&$<$0.07& -12.7&0.2&--& ...\\
627&O9.7&V&-3.7& 50&33600& 600&4.11&0.12&4.11&0.10& -13.0&  0.3&4.67& ...\\
639&O9.7&V&-4.0& 65&33700& 500&4.18&0.10&4.18&0.09& -13.1&  0.2&4.78& SB?\\

\hline
\multicolumn{15}{c}{} \\ [-1.5 ex]
\multicolumn{15}{l}{$^{(1),}$ $^{(2),}$ $^{(3),}$ $^{(4),}$ $^{(5),}$ $^{(6)}$ See corresponding notes to Table~\ref{tab_vz}.} \\
\end{tabular}
\end{table}
\end{landscape}
\begin{table*}[h!!!!!!!!!!!!]
\caption{\label{tab_teff} First estimates of stellar and wind parameters from HHeN analysis of the subsample of stars 
for which \ion{He}{i-ii} analysis alone was considered unreliable (see notes in Sect.~\ref{results}).}
 \centering
\begin{tabular}{crlcccccccl}
\hline
\hline
\multicolumn{11}{c}{} \\ [-2 ex]
VFTS&SpT& LC&$M_v^{(1)}$&$v$\,sin\,$i$&$T_{\rm eff}$&log\,{\em g}& log\,{\em g}$_c^{(3)}$&log\,{\em Q}&$\log{L/L_{\odot}}^{(5)}$ &Comments$^{(6)}$\\
 && &&[$\rm{kms^{-1}}$]&[K]&[dex]&[dex]& &[dex]& \\
\hline
\multicolumn{11}{c}{} \\ [-1.5 ex]
 468&O2&V((f*))+OB&-6.1& 80&52000&4.20&4.20& -12.3&  6.17& VM4 \\
506&ON2&V((n))((f*))&-6.6&100&55000& 4.20&4.20& -12.5&  6.43& SB1s\\
621&O2&V((f*))z&-6.1& 80&54000& 4.20&4.20& -12.7&  6.22& VM3\\
169&O2.5&V(n)((f*))&-5.8&200&47000& 3.90&3.92& -12.5&  5.91& SB?\\
755&O3&Vn((f*))&-5.2&285&46000& 3.90&3.94& -12.7&  5.65& ...\\
797&O3.5&V((n))((fc))z&-5.2&140&45000& 3.80&3.82& -13.0&  5.60& SB?\\
216&O4&V((fc))&-5.9&100&43000&  3.80&3.81& -12.7&  5.83& SB?\\
586&O4&V((n))((fc))z&-4.7&100&45000& 4.00&4.01& -13.0&  5.42& SB?\\
382&O4-5&V((fc))z&-4.8& 75&40000& 3.80&3.81&-13.0&  5.31& ...\\
581&O4-5&V((fc))&-5.0& 70&40000& 3.70&3.71& -12.7&  5.38& ...\\
537&O5&V((fc))z&-4.6& 60&39000& 3.80&3.80& -13.0&  5.19& ...\\
550&O5&V((fc))z&-4.6& 50&39000& 3.80&3.80& -13.0&  5.20& ...\\
577&O6&V((fc))z&-4.4&40&42000&4.00&4.00&-13.0&5.21& ...\\

\hline
\multicolumn{5}{l}{} \\ [-1.5 ex]
\multicolumn{11}{l}{$^{(1),}$ $^{(3),}$ $^{(5),}$ $^{(6)}$ See corresponding notes in Table~\ref{tab_vz}.} \\

\end{tabular}
\end{table*}
 \clearpage
\section{Figures}\label{apendC}

\begin{figure*}[t]
 \centering
  \includegraphics[width=17 cm,angle=0]{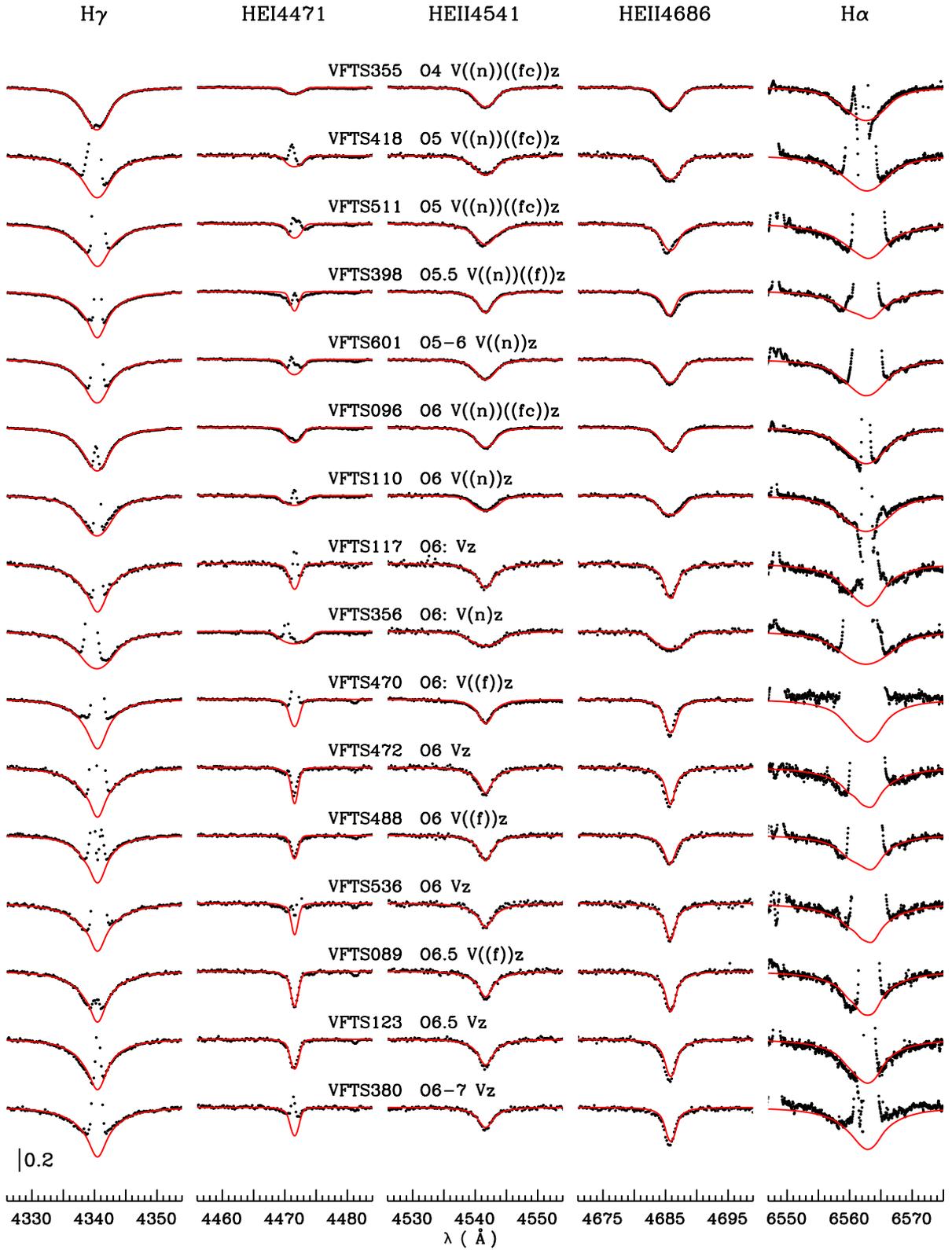}
  \caption{\label{spec_vz1} Observed spectra (black dots) and
    best-fitting synthetic models (red lines) of H$\gamma$,
    \ion{He}{i}\,$\lambda$4471, \ion{He}{ii}\,$\lambda$4541,
    \ion{He}{ii}\,$\lambda$4686 and H$\alpha$ for the O\,Vz subsample.
    Each spectrum is labeled with its VFTS number and spectral
    classification.  The cores of strong nebular lines have been removed.}
  \end{figure*}

\begin{figure*}[t]
 \centering
  \includegraphics[width=17 cm,angle=0]{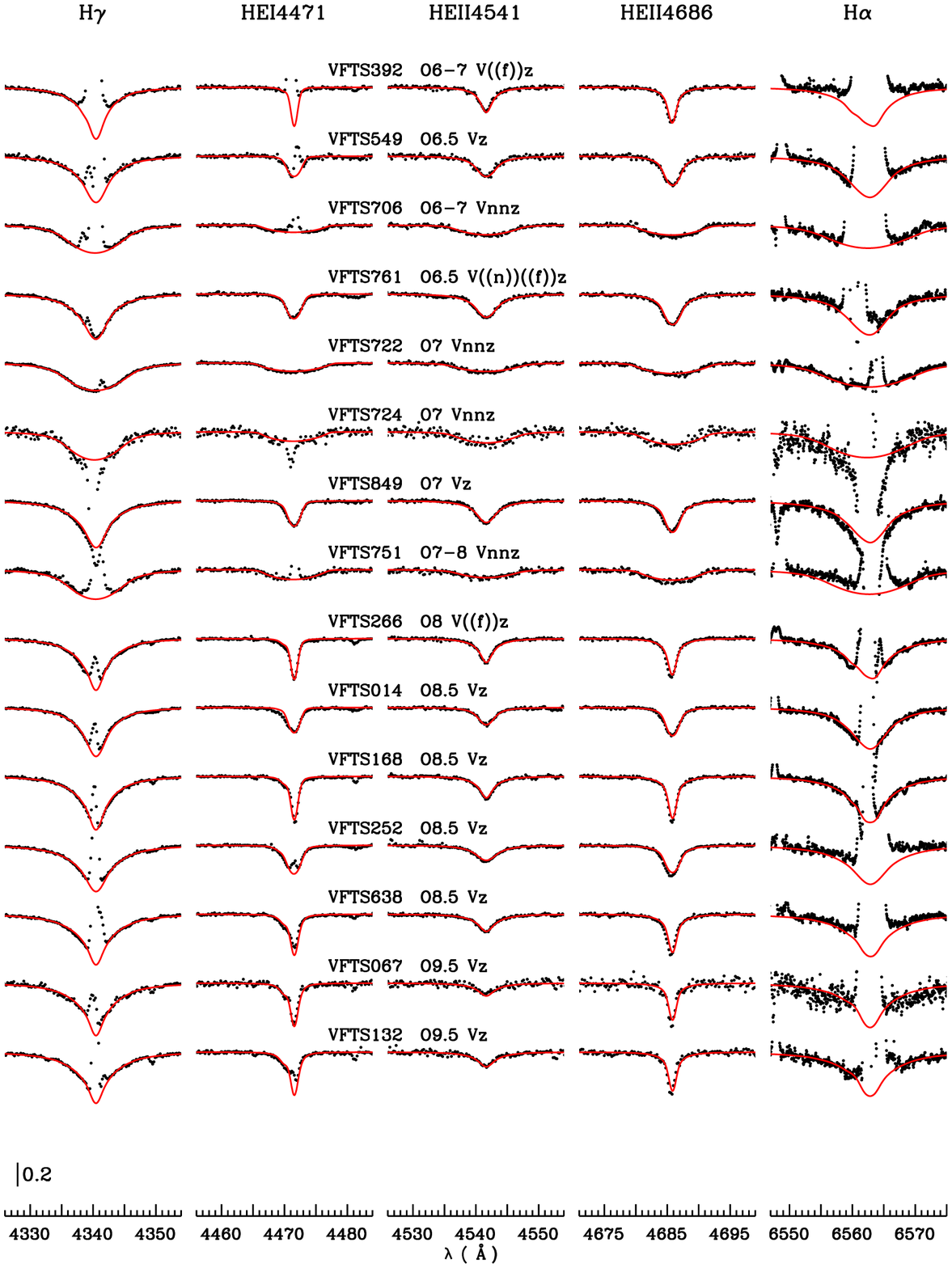}
 \caption{\label{spec_vz2} Same as Fig.~\ref{spec_vz1}.}  
 \end{figure*}

\begin{figure*}[t]
 \centering
  \includegraphics[width=17 cm,angle=0]{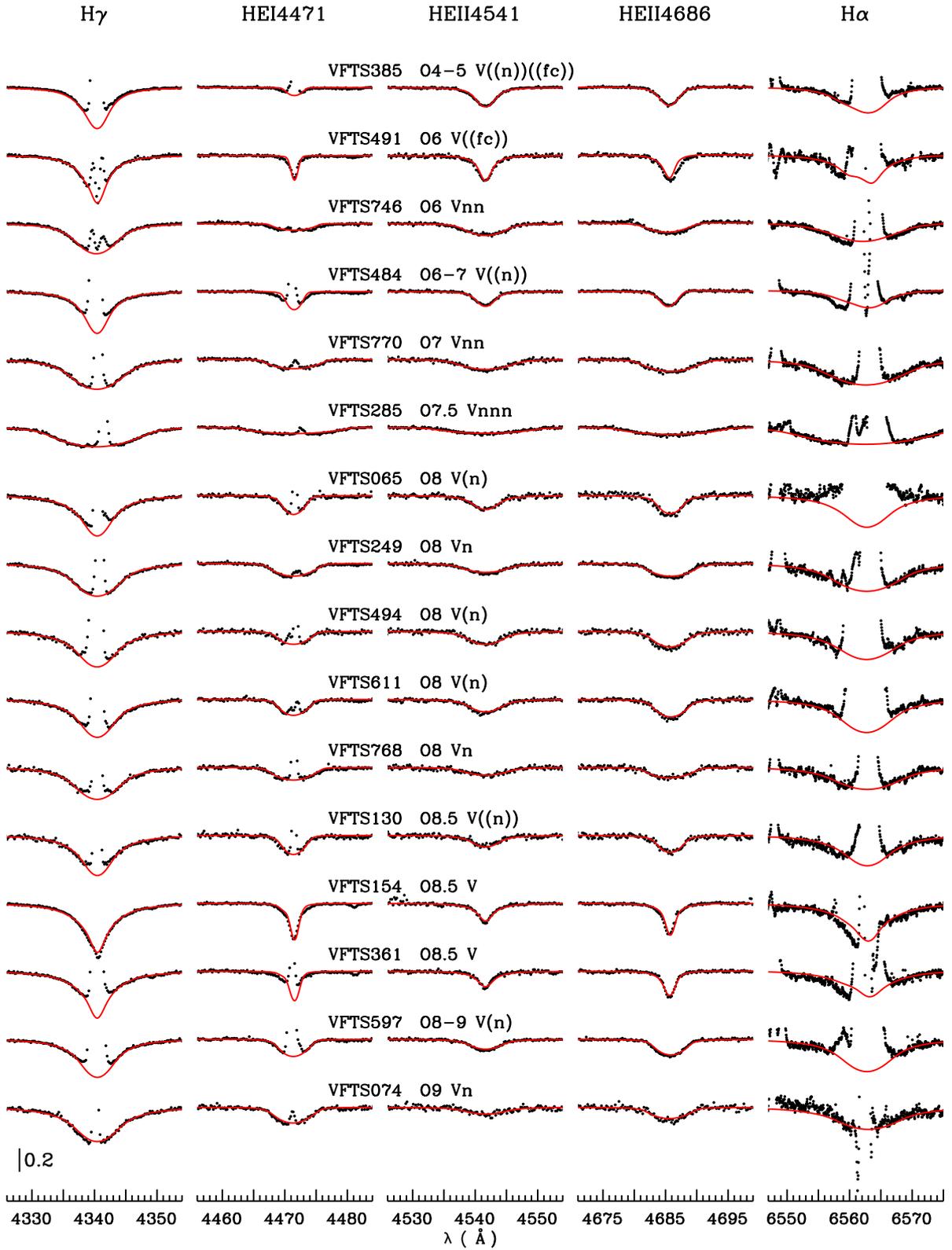}
  \caption{\label{spec_v1} Observed spectra (black dots) and
    best-fitting synthetic models (red lines) for the O\,V\,
    subsample.}
 \end{figure*}

\begin{figure*}[t]
 \centering
  \includegraphics[width=17 cm,angle=0]{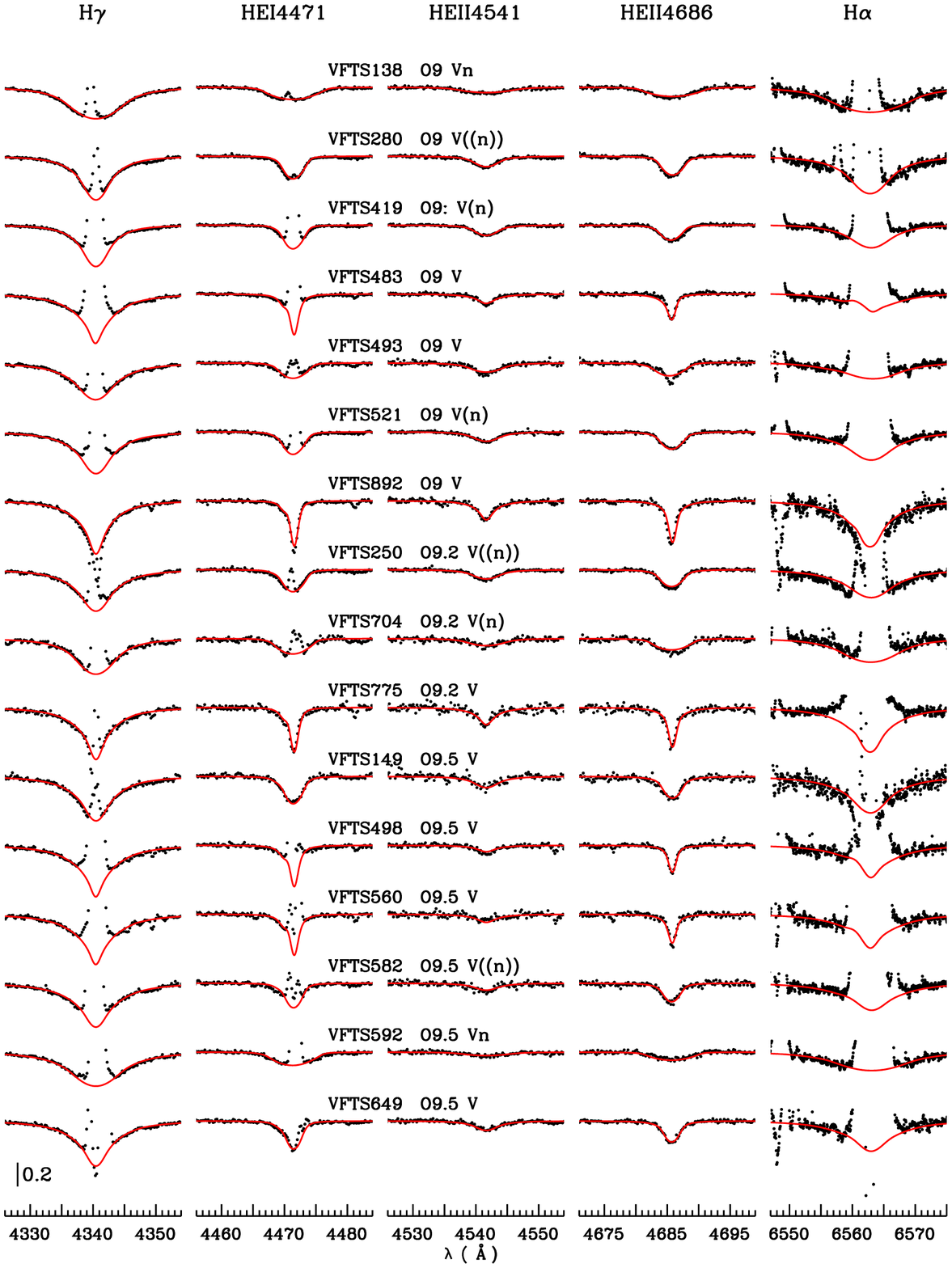}
 \caption{\label{spec_v2} Same as Fig.~\ref{spec_v1}.}  
 \end{figure*}

\begin{figure*}[t]
 \centering
  \includegraphics[width=17 cm,angle=0]{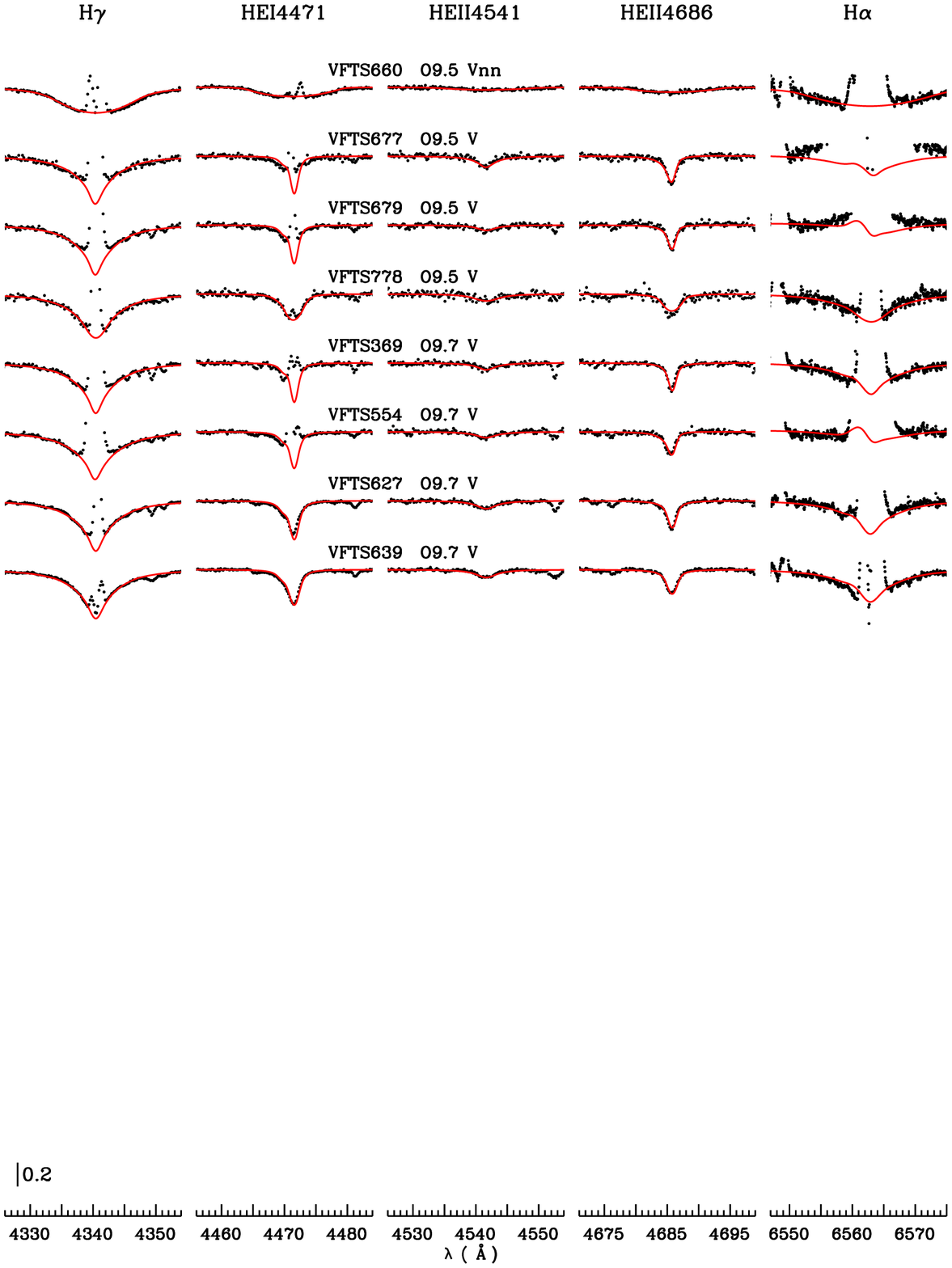}
 \caption{\label{spec_v3} Same as Fig.~\ref{spec_v1}.}  
 \end{figure*}

\begin{figure*}[t]
 \centering
  \includegraphics[width=17 cm,angle=0]{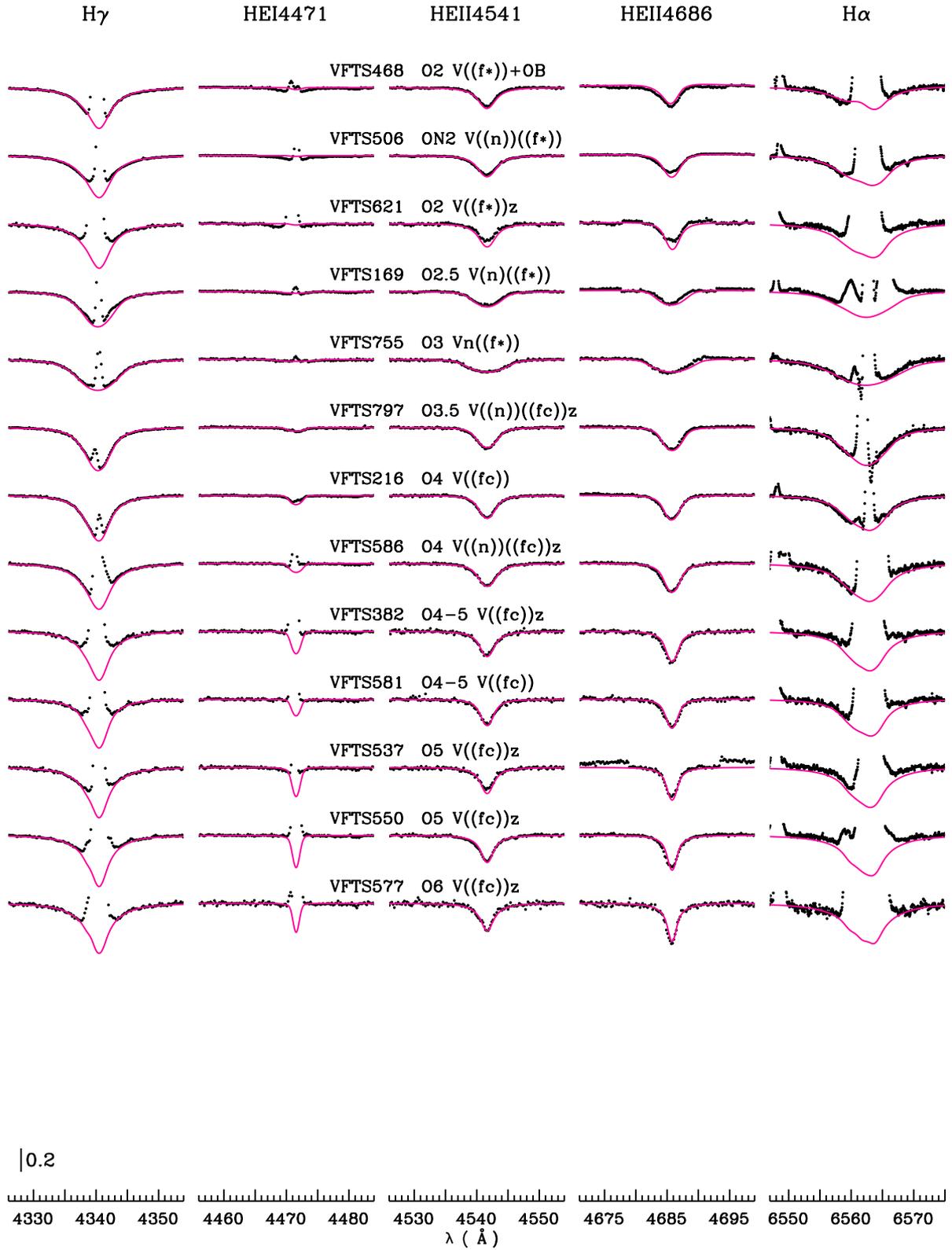}
  \caption{\label{spec_v4} Observed spectra (black dots) and
    best-fitting synthetic models (pink lines) for the O\,Vz and O\,V\
    stars analyzed using H, He and N as the diagnostic lines.}
 \end{figure*}

Figures~\ref{spec_vz1} to \ref{spec_v4} shown the best-fitting models
to the observed VFTS spectra for the most important diagnostic lines
in this study, i.e., H$_{\gamma}$, \ion{He}{i}\,$\lambda$4471,
\ion{He}{ii}\,4541, \ion{He}{i}\,4686, and H$_{\alpha}$.

\end{document}